\documentclass[12pt,preprint]{aastex} %produces a one-column, single-spaced document:

\usepackage{natbib}

\shorttitle{Ab initio EOS and internal structure of Jupiter}
\shortauthors{Nettelmann et al.}

\begin{document} 

\title{Ab initio equation of state data for hydrogen, helium, and water 
        and the internal structure of Jupiter}

\author{Nadine Nettelmann, Bastian Holst, Andr\'e Kietzmann, Martin French, and Ronald Redmer}
\affil{Institut f{\"ur} Physik, Universit{\"a}t Rostock, D-18051 Rostock, Germany}
\email{nadine.nettelmann@uni-rostock.de}

\author{David Blaschke}
\affil{Institute for Theoretical Physics, University of Wroclaw, Max-Born pl. 9, 50-204 Wroclaw, Poland}

\begin{abstract}
The equation of state of hydrogen, helium, and water effects interior structure models of 
giant planets significantly. We present a new equation of state data table, LM-REOS, generated 
by large scale quantum molecular dynamics simulations for hydrogen, helium, and water in the 
warm dense matter regime, i.e.\ for megabar pressures and temperatures of several thousand 
Kelvin, and by advanced chemical methods in the complementary regions. The influence of 
LM-REOS on the structure of Jupiter is investigated and compared with state-of-the-art 
results within a standard three-layer model consistent with astrophysical observations of 
Jupiter. Our new Jupiter models predict an important impact of mixing effects of helium in 
hydrogen with respect to an altered compressibility and immiscibility.
\end{abstract}


%%%%%%%%%%%%%%%%%%%%%%b
\section{Introduction}

Jupiter consists by more than 85\% in mass of hydrogen and helium. 
Among the minor constituents, oxygen contributes the largest fraction. 
Although the major constituents are rather simple elements, they occur almost 
exclusively in extreme, up to now largely unknown states of matter. 
For instance, about 90\% of the planetary material in Jupiter is in a 
high-pressure state beyond the 1~Mbar level up to about 70~Mbar in the center,
and the temperature varies by two orders of magnitude from the cold and dilute 
outer envelope up to about 20.000~K in the deep interior. In terms of usual 
plasma parameters, the interior covers the transition from weak to strong ion 
coupling, and from weak to moderate electron degeneracy without relativistic 
effects. Giant planets like Jupiter are therefore particularly suitable to study 
{\it warm dense matter} (WDM), especially mixtures of H, He, and H$_2$O. 

In this paper we apply for the first time ab initio equation of state (EOS) 
data for H, He, and H$_2$O, representing at least 97\% of the planetary material 
in Jupiter, to interior models in order to derive more reliable implications for an 
improved structure model as well as for the H/He phase diagram.
We have performed large scale quantum molecular dynamics simulations
for the WDM region and applied advanced chemical models in the complementary regions. 
These EOS data for the major components were combined by linear mixing into a new data 
table LM-REOS (Linear Mixing Rostock Equation of State). Fundamental problems concerning 
hydrogen and helium such as the location of the nonmetal-to-metal transition (NMT), 
whether or not this transition is accompanied by a thermodynamic phase instability 
- the plasma phase transition (PPT) -, and the existence of a miscibility gap 
are at the same time key issues for the construction of planetary models.

Strongly improved high-pressure data for planetary materials as well as new and more 
accurate observational data for Jupiter and the other planets have motivated intensive 
efforts to model giant planets. We outline briefly major steps in the development of 
planetary models that lead to our present understanding of Jupiter's interior structure.

%%%% Hubbard 1968,1969
In one of the first approaches, \citet{Hub68} started from estimates of 
Jupiter's net heat flux and inferred a slightly superadiabatic temperature gradient, 
causing convective instability throughout the planet. With his fully homogeneous and, 
thus, overdetermined model he compared with current observational constraints 
in order to discriminate between different equations of state \citep{Hub69}.

%%%% SS 1977, extrapolierte solid state EOS
\citet{SS77-1, SS77-2} examined the transport properties of 
hydrogen-helium mixtures and the influence of a PPT and of H/He phase separation on the 
structure and evolution of hydrogen-helium planets. Either of these processes would 
divide the fluid interior into two convective layers, a molecular outer envolope and a 
metallic inner envelope with different helium abundances. The NMT was predicted to occur 
around 2-4~Mbar, and helium immiscibility between 4-40~Mbar and below 8000~K. Depending 
on the true position of the NMT coexistence line and the demixing curve in the H/He 
phase diagram, the extension of the inhomogeneous, diffusive transition region in between 
could be very narrow. Minor constituents were argued to be redistributed in consequence 
of helium immiscibility.

%%%% SCHL 1992, CSvH 1992 
Voyager data indeed unveiled a depletion of helium in the outer molecular envelope, 
indicating that H/He phase separation has happened there. A thermodynamically consistent 
H-EOS was developed by~\citet{SC92} for astrophysical applications 
within a chemical model that predicted a discontinuous NMT. This PPT has been proposed 
for a long time~\citep{NS1968,NS1970,ZL1943} and the phase diagram was discussed in 
detail~\citep{Beule01,EbNo2003,HNR07}. A first experimental signature for a PPT 
in deuterium was given quite recently~\citep{FortovPRL}. 
Using this H-EOS, the observational constraints for Jupiter could only be adjusted if the 
envelopes were allowed to have a density discontinuity, and if a several Earth masses sized 
core of rocks and ices was admitted~\citep{CSHL92}. For simplicity, helium was 
(and in recent models still is) added to the EOS of hydrogen via the additive volume rule, 
and the fraction of helium in both envelopes was used to adjust the gravitational moments 
$J_2$ and $J_4$. Other constituents than hydrogen and helium - summarized as heavy elements 
or metals - were represented by helium too. 

%%%% Guillot 99, SC95
A significant step forward in constructing reliable Jupiter models was done by
~\citet{Gui99PSS,Gui99} who used an improved computer code~\citep{GM95}, improved EOS data of 
Saumon, Chabrier \& Van Horn~\citeyearpar{SCVH} (SCvH), and more accurate observational data for 
the atmospheric helium abundance from the Galileo mission~\citep{Zahn}. This improved Jupiter 
model predicted a core mass of 0-10~Earth masses ($M_{\oplus}$), a total amount of heavy 
elements of 11-41~$M_{\oplus}$, and a heavy element enrichment of up to 6.5 times the 
solar value in the molecular region. 
The great margins resulted from a variation of the observables within their error bars, 
but most of all from the uncertainty in the H-EOS data itself in the region around the NMT. 
By choosing appropriate pair potentials for the H$_2$ molecules, the underlying H-EOS 
(H-SCvH-ppt) was constructed to reproduce shock compression data derived from gas gun 
experiments~\citep{Nellis83,vanThiel73}. A first-order phase transition was found 
around 2~Mbar, the slightly decreasing coexistence line ends in a second critical point 
at about 15000~K. Its interpolated version (H-SCvH-i) has been widely used in planetary 
modelling~\citep{GZ99,Gui99PSS,SG04}. As before \citep{CSHL92}, the fraction of metals was 
obtained from the excess density with respect to that of a given H/He mixture.

%%%% Kerley, Sesame
In an alternative approach, \citet{Ker04-1} calculated three-layer models 
(a core of rocks or ices and two fluid envelopes) using a revised Sesame EOS for H~\citep{KerleyH}, 
an improved He-EOS~\citep{Ker04-2}, and the PANDA code for the icy components 
H$_2$O, NH$_3$, CH$_4$, and H$_2$S and their atomic constituents to represent metals. 
Choosing threefold enrichment in heavy elements in accordance with measurements~\citep{Atreya03,Mahaffy00} 
for the molecular envelope, an enrichment of 7.5 times solar was required in the metallic envelope 
to match $J_2$, in total 35~$M_{\oplus}$. The large heavy element enrichment in the inner envelope 
was argued to result from a dispersion of planetesimals during the formation process. 

%%%% Saumon Guillot 2004
The latest extensive calculations of Jupiter models have been performed \citep{SG04} using 
the Sesame EOS 7154 of water and the Sesame EOS 7100 of dry sand to represent metals within 
the envelopes and for hydrogen using the simple linear mixing (LM) model of Ross~\citep{HRossN95}. 
The LM model served as a basis for the generation of different H-EOS representing the current uncertainty 
on the experimental deuterium EOS data. Not all of the H-EOS applied gave satisfactory results. 
While acceptable Jupiter models were placed within the range of the previous models cited above, 
some LM-variants as well as the original Sesame EOS 5263 were not compatible with Jupiter's gravitational 
properties or age. 

%%%% other EOS, people
Other attempts~\citep{HNR07} to construct Jupiter models with the Sesame EOS of hydrogen or a combination of 
fluid variational theory with plasma effects (FVT$^+$)~\citep{HNR07} also failed to reproduce the 
observational constraints.

%%%% Experimente
To decide whether a mismatch between calculated and observed parameters is due to an inaccuracy 
of the EOS or due to improper assumptions in the planetary model, accurate experimental data for the 
high-pressure phase diagram of hydrogen and hydrogen-helium mixtures are much-needed. Various 
experiments have been performed in the recent years to study the complex physical behavior of 
hydrogen in the region up to several Mbar~\citep{Boriskov05,Collins98,Knudson,Weir+96}. The 
transition from a nonconducting, molecular fluid to an atomic, conducting fluid can be derived 
from electrical conductivity~\citep{Nellis96} as well as reflectivity data~\citep{Celliers00}. 
For instance, the NMT in fluid hydrogen at temperatures of about 3000~K has been pinpointed 
at 1.4~Mbar in multiple shock wave experiments~\citep{Weir+96}. Reflectivity measurements 
indicate an NMT along the Hugoniot curve at about 0.5~Mbar~\citep{Celliers00}. Most recently, 
the first experimental signature of a first-order phase transition connected with this NMT 
has been found in quasi-isentropic shock compression experiments in deuterium~\citep{FortovPRL}.
The maximum compression ratio along the principal Hugoniot curve of hydrogen is not ultimately fixed 
but a value of 4.25 has been found in various shock-wave experiments~\citep{Boriskov05,Knudson} 
which is not predicted by any of the theoretical H-EOS mentioned above except the Sesame EOS. 

%%%% ab initio EOS
On the other hand, current ab initio calculations yield not only agreement with experimental 
Hugoniot curves ~\citep{Len+00,Des03,Bon+04,Hol+08} but also with reflectivity data~\citep{Col+01,Maz+03,Hol+08} 
at high pressures. Therefore, we use in this study ab initio EOS data derived from extensive 
quantum molecular dynamics (QMD) simulations for the most abundant planetary materials 
hydrogen~\citep{Hol+08}, helium~\citep{Kie+07}, and water~\citep{Fre+08} and study the impact 
of these data on models of Jupiter within the three-layer structure assumption. 

%%% Paper-Ouline
Our paper is organized as follows. In the following section we describe our procedure to calculate 
three-layer planetary models. In \S\ref{chap_EOS} the new LM-REOS data tables for H, He, and H$_2$O 
are introduced and the differences to the widely used SCvH EOS are exemplified by means of the Jupiter 
isentrope and the relevant plasma parameters. Our results for Jupiter are presented in 
\S\ref{chap_resJup}. First, our numerical results are checked by using the SCvH EOS. We then process  
the new LM-REOS data tables and compare the predictions for Jupiter's interior with previous 
results~\citep{Gui99PSS, Gui03} and with additional calculations using the SCvH-ppt EOS regarding the 
distribution of chemical species. 
We discuss our results for Jupiter in \S\ref{chap_discussion} and propose key issues for future work 
on high-pressure EOS data and improvements of planetary models. 
A summary is given in \S\ref{chap_summary}.

%%%%%%%%%%%%%%%%%%%%%%%%%%%%%%%%%%%%%%%%%%%%%%%%%%%%%
\section{Three-layer planetary models}\label{chap_planIntMods}

%%%%%%%%%%%%%%%%%%%%%%%%%%%%%%%%%%%%%%%%
\subsection{Parameters for Jupiter}

Our interior models follow the approach described in detail by~\citet{Gui99PSS}. 
The planet is assumed to consist of three layers: two fluid, homogeneous, and isentropic envelopes 
composed of hydrogen, helium and metals and an isothermal core of rocks or ices. 

For given EOS of these materials, the precise fractions of helium and metals within the envelopes 
and the size of the core are constrained by the requirement of matching the observational constraints, 
see Tab.~\ref{tab_obs}. Jupiter's mass-radius relation (total mass $M_{\rm J}$, equatorial radius $R_{eq}$) 
requires the dominant mass fraction to be attributed to hydrogen~\citep{Guillot05}. The mean helium fraction 
$\bar{Y}$ is assumed to reflect the hydrogen-helium ratio of the protosolar cloud. Due to the assumption 
of homogeneous layers the atmospheric helium fraction $Y_{\rm mol}$ is assigned to the whole outer envelope, 
starting at the 1~bar level with temperature $T_1$. The adjustment of the gravitational moments $J_2$ and $J_4$ 
assuming rigid rotation with angular velocity $\omega$ sets limits to the fraction of metals $Z_{\rm mol}$ 
in the outer envelope and $Z_{\rm met}$ in the inner envelope. Observational error bars translate into uncertainties 
of the resulting values of these structure properties, allowing for a variety of {\it acceptable} models. 
Comparing acceptable models with other predictions, such as from formation theory or abundance measurements, 
allows to evaluate the models, see \S\ref{chap_discussion}.

\placetable{tab_obs}
\clearpage
\begin{deluxetable}{cc}
%\setlength{\footnotesep}{1pt}
%\tablewidth{}
%\tablecolumns{}
\tablecaption{\label{tab_obs}Measured data that are considered in the modelling procedure.}
\tablehead{
\colhead{Parameter} & \colhead{Jupiter \tablenotemark{a}}}
\startdata
mass $M_{\rm J}$ [g]					& $1.8986112(15)\times 10^{30}$\\
equatorial radius $R_{eq}$ [m]	& $7.1492(4)\times 10^7$ \tablenotemark{b}\\
average helium mass fraction $\bar{Y}$	& $0.275^{+0.003}_{-0.005}$ \tablenotemark{c}\\
helium mass fraction $Y_{\rm mol}$	&	0.238(7)		\tablenotemark{d}\\
temperature $T_1$ [K]					& 165-170					\tablenotemark{e}\\
angular velocity $\omega$			& 2$\pi$/9h55 			\tablenotemark{b}\\
$J_2 / 10^{-2}$							& 1.4697(1) 			\tablenotemark{b}\\
$J_4 / 10^{-4}$							& -5.84(5) 				\tablenotemark{b}\\
\enddata
\tablenotetext{a}{Numbers in parenthesis are the uncertainty in the last digits of the given value.}
\tablenotetext{b}{taken from~\citet{Gui03}.}
\tablenotetext{c}{\citet{Bahcall}.}
\tablenotetext{d}{\citet{Zahn}.}
\tablenotetext{e}{taken from~\citet{Gui99PSS}}
\end{deluxetable}

\clearpage

Besides being the simplest interior structure model that is consistent with the observational 
constraints mentioned above, the three-layer stucture has been discussed and evaluated in connection with 
various underlying or neglected physical properties~\citep{Gui99PSS}. For instance, a small opacity could occur 
in a narrow region around 1500~K~\citep{GGCM94, GCMG94}, causing a stable radiative layer and thus lengthen 
the cooling time of the planet without affecting the element distribution. Differential rotation is observed 
in the atmosphere and could modify $J_2$ up to 0.5\%  and $J_4$ up to 1\% \citep{ZT}. Assuming more than two 
envelopes would probably reflect the real structure of Jupiter more adequately, but also entail more degrees 
of freedom than can be adjusted by the constraints presently available.

Layer boundaries that devide the fluid planet into convective layers with different compositions 
have been predicted theoretically from general thermodynamic properties of warm dense hydrogen and 
hydrogen-helium mixtures~\citep{SS77-1, SS77-2}. Possible sources of inhomogeneity from thermodynamic considerations 
are phase separation due to first-order phase transitions or immiscibility between different components, 
and from evolution theory the presence of a slowly eroding core. In case of Jupiter, the PPT in hydrogen 
is a candidate for an important first-order phase transition, besides those for the minor constituents as, 
e.g., H$_2$O in the outermost shells. Phase separation may occur between neutral helium and metallic hydrogen 
near the PPT of hydrogen~\citep{Klepeis, Pfaff}. Both effects motivate a three-layer structure assumption.

%%%%%%%%%%%%%%%%%%%%%%%%%%%%%%%%
\subsection{Basic equations}\label{chap_bequs}

The construction of acceptable interior models is performed in an iterative procedure. 
Each model is defined by a set of distinct parameter values to be matched. For the 
accurately known $M_{\rm J}, R_{eq}, \omega, J_2$, these values are the observed ones within 
an error of one standard deviation ($1\sigma$); for the less accurately known $\bar{Y}$ and $J_4$, 
the values are chosen in the $1\sigma$ range with a computational error of less than 10\% of their 
$1\sigma$ error. The interior models are not optimized to meet $J_6$. Resulting values of $J_6$ lie 
always in the $1\sigma$ range, if $J_2$ and $J_4$ do. For a H-EOS with PPT, the transition pressure 
$P_{\rm m}$ from the molecular to the metallic layer is chosen to coincide with the entry of the outer 
isentrope into the instability region. For a H-EOS without PPT, the pressure is varied between 1.4 
and 5~Mbar. For a distinct choice of mass fractions $X$ of hydrogen, $Y$ of helium, $Z$ of metals, 
the EOS of the mixture is calculated, see \S\ref{chap_mixtEOS}. From these tabulated thermal and 
caloric EOS of the mixture, the entropy can be calculated via thermodynamic relations. 
The isentropes of the outer and of the inner envelope and the isotherm of the core are defined by 
the boundary conditions $(T_1,P_1)$, $(T_{\rm m},P_{\rm m})$, and $(T_{\rm c}, P_{\rm c})$, where the 
transition temperature $T_{\rm m}$ results from the outer isentrope at $P=P_{\rm m}$, and the core 
temperature $T_{\rm c}$ from the inner isentrope at $P=P_{\rm c}$. Since the size of the core is not 
known in advance, $P_{\rm c}$ is varied within the iteration scheme described below until the total mass 
condition holds. Along this curve of piecewise constant entropy or temperature the equation of motion 
describing hydrostatic equilibrium, 
\begin{equation}\label{eq_motion}
	\rho^{-1}\nabla P = \nabla (V+Q)\:,
\end{equation}
is integrated inwards, where $\rho$ is the mass density and $U=V+Q$ the total potential composed of 
the gravitational potential 
\begin{equation}\label{eq_gravipotAllg}
	V(\vec{r})= G \int\! d^3r'\,\frac{\rho(\vec{r}\,')}{|\vec{r}\,'-\vec{r}|}
\end{equation}
and the centrifugal potential $Q$ defined by 
\begin{equation}\label{eq_centripotAllg}
	-\nabla Q(\vec{r}) = \vec{\omega}\times(\vec{\omega}\times\vec{r})\:.
\end{equation}
A proper description for the potential of axisymmetric but, due to rotation, oblate planets is a 
multipole expansion, which transforms the general expressions (\ref{eq_gravipotAllg}) and 
(\ref{eq_centripotAllg}) into the following forms with spherical coordinates $r,\theta$, and 
Legendre polynomials $P_{2n}(t=\cos\theta)$
\begin{equation}\label{eq_gravipotMulti}
	V(r,\theta) = \frac{G}{r}\sum_{n=0}^{\infty}P_{2n}(t) \int\! d^3r'\,\rho(r',\theta')
	\left(\frac{r'}{r}\right)^{2k}P_{2n}(t')\:,
\end{equation}
\begin{equation}
	Q(r,\theta) = \frac{1}{3}\omega^2r^2\left(1-P_2(t)\right)\:.
\end{equation}
The index $k=n$ describes the external gravitational potential for $r>r'$ and $k=-(n+1)$ the internal 
gravitational potential for $r<r'$. At the surface, the internal gravitational potential vanishes and 
$V$ reduces to
\begin{equation}
	V(r,\theta)	=	\frac{GM}{r}\left(1-\sum_{n=0}^{\infty}\left(\frac{R_{eq}}{r}\right)^{2n}
	J_{2n}P_{2n}(t)\right) \:,
\end{equation}
with the gravitational moments
\begin{equation}\label{eq_J2n}
	J_{2n}=\frac{1}{MR_{eq}^{2n}}\int\! d^3r'\,\rho(r',\theta')r'^{2n}P_{2n}(t')\:.
\end{equation}
Calculating the integrals in equations (\ref{eq_gravipotMulti}) and (\ref{eq_J2n}) requires knowledge 
of the planetary shape and the density distribution. We apply the \textit{Theory of Figures}~\citep{ZT} 
and the method described therein and in \citep{NRB07} for solving the system of integral equations 
on computers to iteratively calculate the planetary shape, the potential, and the density distribution. 
The key point of this theory is to reduce the dimension of the problem by replacing the former radial 
coordinate $r$ by its representation 
\begin{equation}\label{eq_r2l}
	r(\theta)=l\left(1+\sum_{i=0}^\infty s_{2i}(l)P_{2i}(t)\right)
\end{equation}
on equipotential surfaces ($l=$const) with expansion coefficients $s_{2i}(l)$. Reasonable and practical 
cutoff indices for the sum in equation~(\ref{eq_r2l}) are 3, 4 or 5. In accordance with the treatment 
of~\citet{Gui99PSS}, we consider only coefficients up to third order, i.e.\ $i=3$, which introduces 
an error into the calculation of the gravitational moments of the order of magnitude of $J_6$ or about 
five times the observational error of $J_2$. In terms of the level coordinate $l=0\ldots\bar{R},$ 
the equations to be integrated along the piecewise isentrope take the simple form
\begin{equation}\label{eq_dPdl}
	\frac{1}{\rho(l)}\frac{dP(l)}{dl} = \frac{dU(l)}{dl}
\end{equation}
for the equation of motion and
\begin{equation}\label{eq_dmdl}
	\frac{dm}{dl}=4\pi l^2\rho(l)
\end{equation}
for the equation of mass conservation. The mean radius $\bar{R}$ of the surface (at the 1-bar level) is 
not known in advance and has to be adjusted in order to satisfy $r(\theta=\pi/2)=R_{eq}$. The 
coefficients $s_{2i}$ in equation~(\ref{eq_r2l}), the \textit{figure functions}, determine the shape of 
the rotating planet. As suggested by~\citet{ZT} they are calculated for a fixed density distribution 
$\rho(l)$ in an iterative procedure until convergence of at least 0.01\% for $s_2$ is reached, which 
turns out to be accompanied by an error of 0.05\% for $s_4$ and 0.1\% for $s_6$. The converged figure 
functions enter into equations~(\ref{eq_dPdl}), (\ref{eq_dmdl}), and (\ref{eq_J2n}) and a revised 
density distribution is calculated resulting in a different core mass and, thus, a different central 
pressure $P_{\rm c}$. After about six iterations of calculating converged figure functions and 
corresponding density distributions, $J_2$ and $J_4$ are converged within their observational 
error bars, but not necessarily to the observed values $J_2^{\rm obs}$, $J_4^{\rm obs}$. To achieve 
agreement also with $\bar{Y}$, the guess of $Z_{\rm mol}$, $Z_{\rm met}$ and $Y_{\rm met}$ is improved 
slightly and the procedure decribed above is repeated, starting with the present density distribution, 
until the desired values of $J_2$, $J_4$ and $\bar{Y}$ have been approached. This procedure can, 
in principle, be applied to all giant planets in the Solar system.

%%%%%%%%%%%%%%%%%%%%%
\subsection{Accuracy}

In order to give reliable results for the interior structure model and to study the influence of different 
EOS data sets, the numerical treatment has to ensure a definite precision. The accuracy of acceptable 
interior models should be in the order of or smaller than the relative error of the most accurate observed quantity 
to be adjusted during the procedure. For M$_{\rm J}$, R$_{eq}$ and $J_2$ the observational error is below 0.01\% 
and for $\bar{Y}$, $Y_{\rm mol}$, $T_1$, and $J_4$ larger than 1\%. The uncertainty associated with the QMD data 
themselves induces an error of the order of 1\% to the isentropes. Hence we consider a numerical accuracy of 0.1\% 
as sufficient to study the effect of the uncertainty of the less accurately known observables and for conclusive 
interior models regarding the EOS applied.

Numerical errors occur due to the integration of equations (\ref{eq_dPdl}) and (\ref{eq_dmdl}), the differentiation 
of the gravitational potential, and the integrals in equation (\ref{eq_gravipotMulti}), transformed to the level 
coordinate $l$. We have tested the accuracy of integrating the differential equations (\ref{eq_dPdl}) and (\ref{eq_dmdl}) 
for a non-rotating polytrope of index 1. Our numerical results for the profiles $m(r)$ and $P(r)$ obtained via a 
fourth-order Runge-Kutta method with adaptive stepsize control in regions with steep gradients differ by less than 
0.001\% from the analytical solution in the outer 80\% of the planet. In the inner 20\% of the planet including the 
core region, a shooting splitting method is applied to localize the outer boundary of the core. We then search for 
a solution starting from the center to meet the envelope solution there with a deviation of less than 0.1\%. 
Since a polytropic model does not have a core, this method cannot be applied there and the difference rises to 50\% 
for the mass and 0.1\% for the pressure near the center.

To estimate the error resulting from the calculation of the integrals and its enhancement during the iteration procedure, 
we have compared fully converged planetary models. The resulting mass fractions $Z_{\rm mol}$ and $Z_{\rm met}$ 
and the size of the core differ by less than 2\% if the number of intervals $dl'$ in the integrations is doubled. 
Since density discontinuities at layer boundaries are extended over a small but finite intervall, they do not cause 
numerical difficulties. Extending this intervall from 0.001 $R_{\rm J}$ to 0.01 $R_{\rm J}$ effects the resulting 
core mass by about 0.1 $M_{\oplus}$ and the heavy element abundance by about $2\%$.

We conclude that a convergence of our numerical procedure within 0.1\% can be ensured, 
but the resulting values for the model parameters are uncertain within $\pm 0.1$ M$_{\oplus}$ or 2\%.

%%%%%%%%%%%%%%%%%%%%%%%%%%%%%%%%%%%%%%%%%%%%
\section{EOS data}\label{chap_EOS}

Interior models of Jupiter's present state require EOS data for H, He, and metals from about 160~K at 1~bar up to 
25000~K at about 45~Mbar, and for core materials (rocks, ices) of 35-80~Mbar around 20000~K. No state-of-the-art 
thermodynamic model is capable of calculating accurate EOS data for all planetary materials in this wide 
pressure-temperature range. However, various EOS data sets have been constructed for this purpose by adopting 
known limiting cases and interpolating in between or extending them till phase boundaries. Examples are the 
Sesame EOS data tables 5251, 5761 and 7150 for H, He and H$_2$O~\citep{SesameDatabase} and chemical models 
such as the SCvH EOS~\citep{SCVH} and FVT$^+$~\citep{HNR07}. These data sets have the largest uncertainties 
in the WDM regime. We have, therefore, performed extensive QMD simulations to generate accurate EOS data for 
warm dense H, He and H$_2$O, i.e.\ for moderate temperatures and densities higher than 0.2 (H, He) or 
1~g~cm$^{-3}$ (H$_2$O). The new ab initio data cover at least 97\% of the planetary material inside Jupiter.
For lower densities, the simulation times increases dramatically so that other methods have to be applied.

We outline the method of QMD simulations below and describe the construction of improved EOS tables for H, He, 
and H$_2$O by combining QMD results for WDM with chemical models in complementary regions of the phase diagram. 
Predictions of these new EOS tables are compared with SCvH EOS with respect to the Jupiter adiabat and plasma parameters.

%%%%%%%%%%%%%%%%%%%%%%%%%%%%
\subsection{QMD simulations}

For the QMD simulations we have used the code VASP (Vienna Ab Initio Simulation Package) developed 
by \cite{VASP1,VASP2} and \citet{VASP3}. Within this method, the ions are treated by classical 
molecular dynamics simulations. For a given ion distribution, the density distribution of the 
electrons is determined using finite temperature density functional theory (FT-DFT). The electronic 
wave functions are represented by plane waves and calculated using projector augmented wave (PAW) 
pseudopotentials \citep{PAW_B,PAW_KJ,Desj02}. The central input into DFT is the exchange-correlation 
functional accounting for interactions between electrons and ions and the electrons themselves. 
It is calculated within generalized gradient approximation (GGA) using the parametrization of~\citet{PBE}. 

% convergence
The convergence of the thermodynamic quantities in QMD simulations is an important issue. We found 
that a plane wave cutoff energy of 700 (He), 900 (H$_2$O, see \citet{MD06}), and up to 1200~eV 
(H, see~\citet{Des03}) is necessary to converge the pressure within 3\% accuracy. Furthermore, we 
have checked the convergence with respect to a systematic enlargement of the $\textbf{k}$-point set 
in the representation of the Brillouin zone. Higher-order $\textbf{k}$ points modify the EOS data 
only within 1\% relative to a one-point result. Therefore, we have restricted our calculations to 
the $\Gamma$ point for water or the mean value point $(1/4, 1/4, 1/4)$~\citep{Baldereschi} for H and He. 
The overall uncertainty of the QMD data due to statistical and systematic errors is below 5\%. 
% T,V,N
The simulations were performed for a canonical ensemble where the temperature, the volume of the 
simulation box, and the particle number in the box are conserved quantities. We consider 32 to 
162~atoms in the simulation box and periodic boundary conditions. The ion temperature $T_i$ is 
regulated by a Nos\'{e}-Hoover thermostat \citep{Nose84} and the electron temperature 
(with $T_e=T_i$) is fixed by Fermi weighting the occupation of bands using Mermins approach~\citep{Mermin65}. 
% appropriate T-range
At high temperatures, the number of bands to be considered increases exponentially with the electron 
temperature so that a treatment of high-temperature plasmas is rather challenging. However, 
simulations of WDM states at several 1000~K as typical for planetary interiors are performed 
without any serious difficulties. When the simulation has reached thermodynamic equilibrium, 
the subsequent 500 to 2000~time steps are taken to calculate the EOS data as running averages.

%%%%%%%%%%%%%%%%%%%%%%%%%%%%%%%%%%%%%%%%%%%%%
\subsection{Construction of the hydrogen EOS}

For densities below 0.1~g~cm$^{-3}$ we have used FVT$^+$ which treats the dissociation of molecules 
self-consistently and takes into account the ionization of the hydrogen atoms via a respective mass 
action law~\citep{HNR07}. FVT$^+$ makes use of Pad{\'e} formulas \citep{ChabPot} for the plasma properties. 
Below 3000~K ionization does not play a role at planetary pressures and is thus neglected. 
FVT$^+$ data on the one hand and QMD data on the other hand are then combined within a spline interpolation. 

\begin{figure}
\plotone{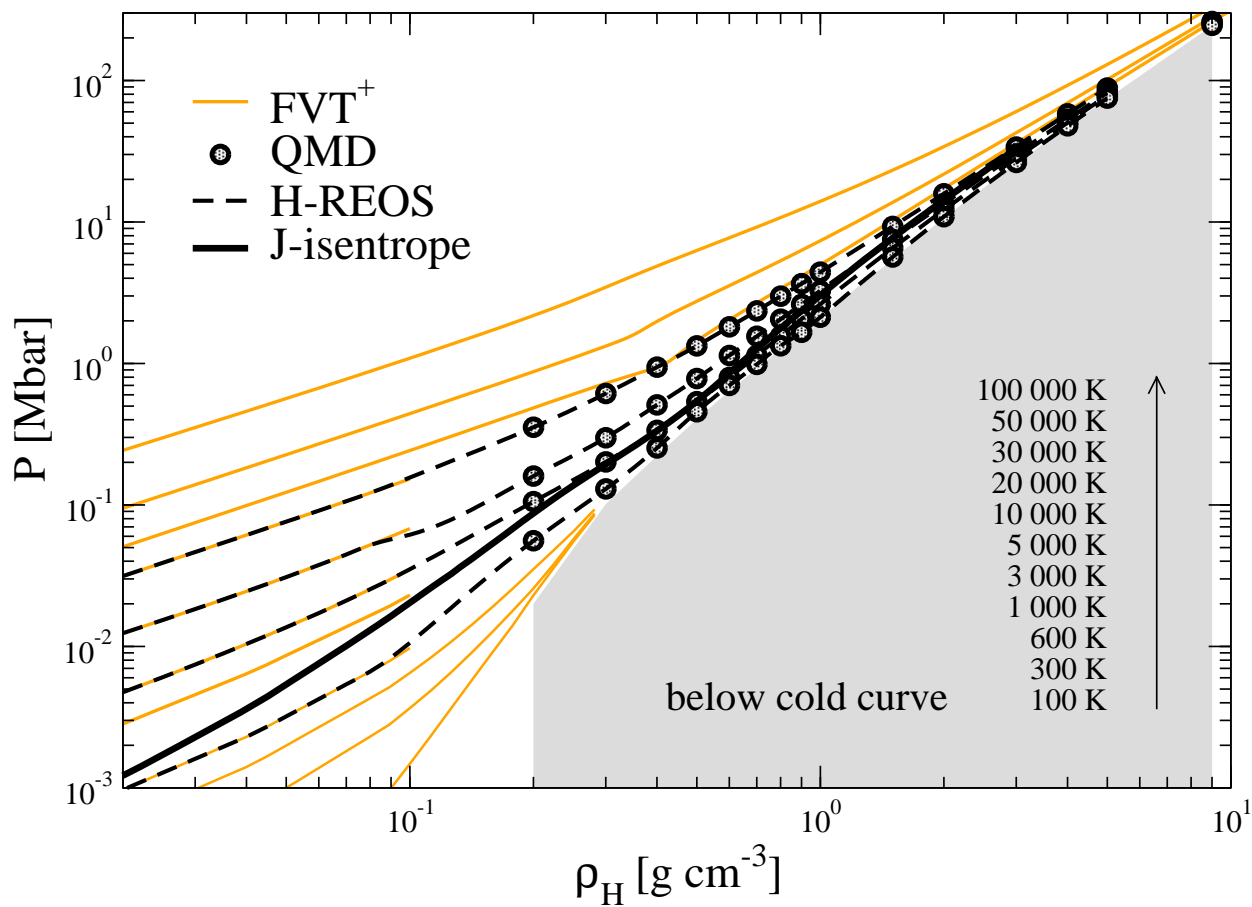}
%\plotone{./f1_nocolor.eps}
\figcaption{\label{fig_H_isoTs}(Color online) Isotherms for hydrogen for different EOS tables: FVT$^+$, QMD, and H-REOS. The thick line shows the pressure-density relation of the hydrogen fraction of a 
typical Jupiter isentrope and the grey area masks the non-accessible region below the cold-curve represented here by a 100 K isotherm of QMD data.}
\end{figure}

Both methods yield different reference values $U_0$ for the internal energy, which have been fixed by the 
Hugoniot starting point of fluid, molecular hydrogen at 19.6~K and 0.0855 g~cm$^{-3}$. The energy difference 
is added to the QMD data in order to get a smooth transition between the data sets. The final hydrogen EOS 
is referred to as \textsl{H-REOS}. Examples of isotherms are shown in Fig.~\ref{fig_H_isoTs}.

\placefigure{fig_H_isoTs}%f1.eps

%%%%%%%%%%%%%%%%%%%%%%%%%%%%%%%%%%%%%%%%%%%
\subsection{Construction of the helium EOS}

\begin{figure}
\plotone{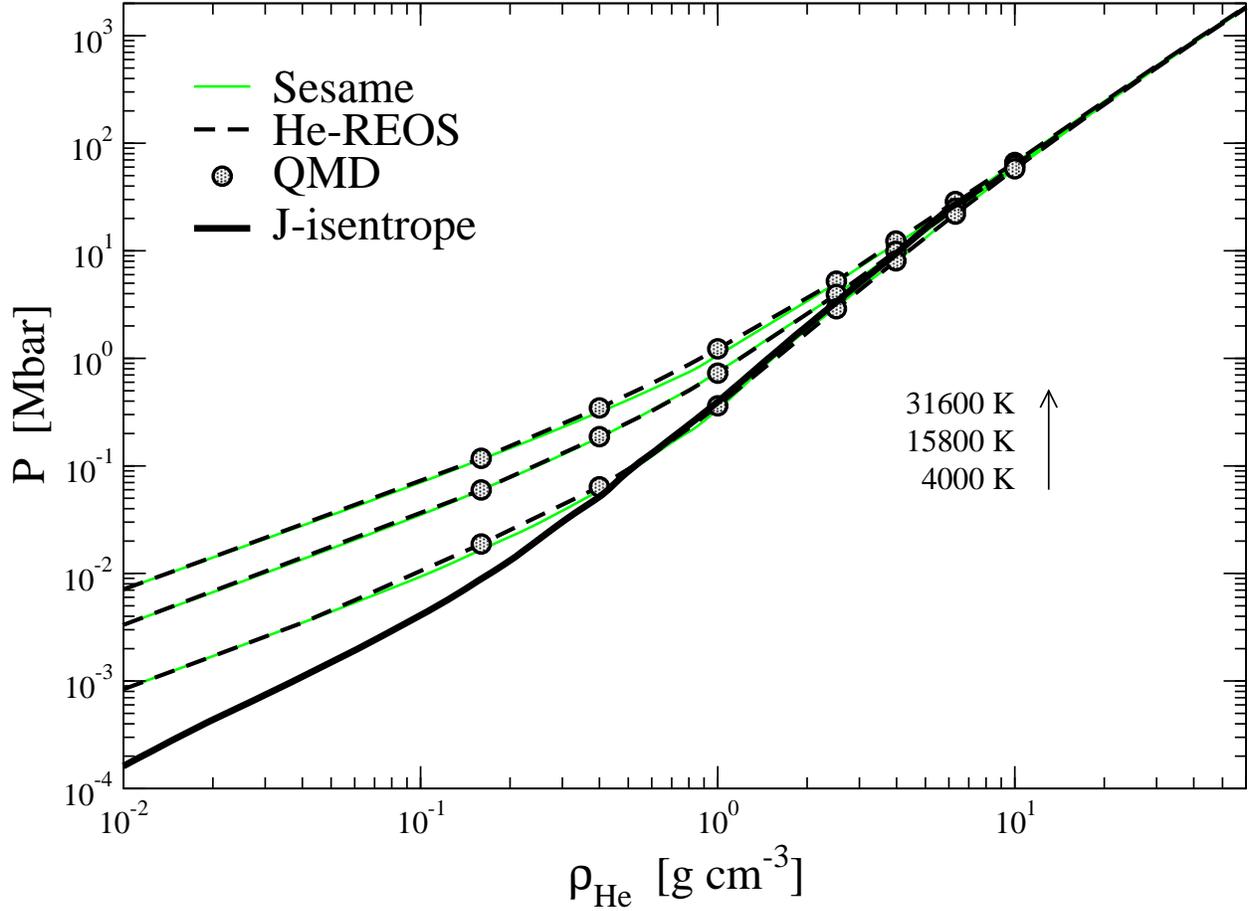}
%\plotone{./f2_nocolor.eps}
\figcaption{\label{fig_He_isoTs}(Color online) Isotherms for helium for T=4000, 15800, 31600~K and three different EOS tables: Sesame 5761, QMD data, and He-REOS. The thick line shows the pressure-density relation of the helium fraction along a typical Jupiter isentrope.} 
\end{figure}

The He-EOS named \textsl{He-REOS} is a combination of QMD data for densities between 
0.16 and 10~g~cm$^{-3}$ and temperatures between 4000 and 31600~K, and Sesame 5761 
data~\citep{SesameDatabase} for densities below 0.01 and above 16~g~cm$^{-3}$ and for 
lower and higher temperatures. For intermediate densities, the isotherms of both data sets 
were interpolated. In Fig.~\ref{fig_He_isoTs} we compare isotherms of He-REOS with Sesame 5761. 
QMD data for helium indicate a slightly lower compressibility around 1~Mbar. Optical properties, 
especially the metallization of fluid helium, as relevant for giant planets and white dwarf atmospheres 
have been investigated recently \citep{Kie+07,KMSC07} by means of QMD simulations. 

\placefigure{fig_He_isoTs}%f2.eps

%%%%%%%%%%%%%%%%%%%%%%%%%%%%%%%%%%%%%%%%%%
\subsection{Construction of the water EOS}

The EOS of water named \textsl{H$_2$O-REOS} is a combination of four EOS data sets. 
QMD data are considered for 1000 $\leq T \leq$ 10000~K and between 2 and 7~g~cm$^{-3}$, 
as well as for 10000 $\leq T\leq$ 24000~K and between 5 and 15~g~cm$^{-3}$. 
Details of the QMD simulations for H$_2$O and the respective EOS data will be published 
elsewhere~\citep{Fre+08}. For the phases ice~I and liquid water we apply accurate 
data tables named \textsl{FW}~\citep{FWeis} and \textsl{WP}~\citep{WPwasser}, respectively. 
All other regions are filled up with Sesame 7150 data~\citep{SesameDatabase}. 
Intermediate EOS data are gained via interpolation. 

\begin{figure}
\plotone{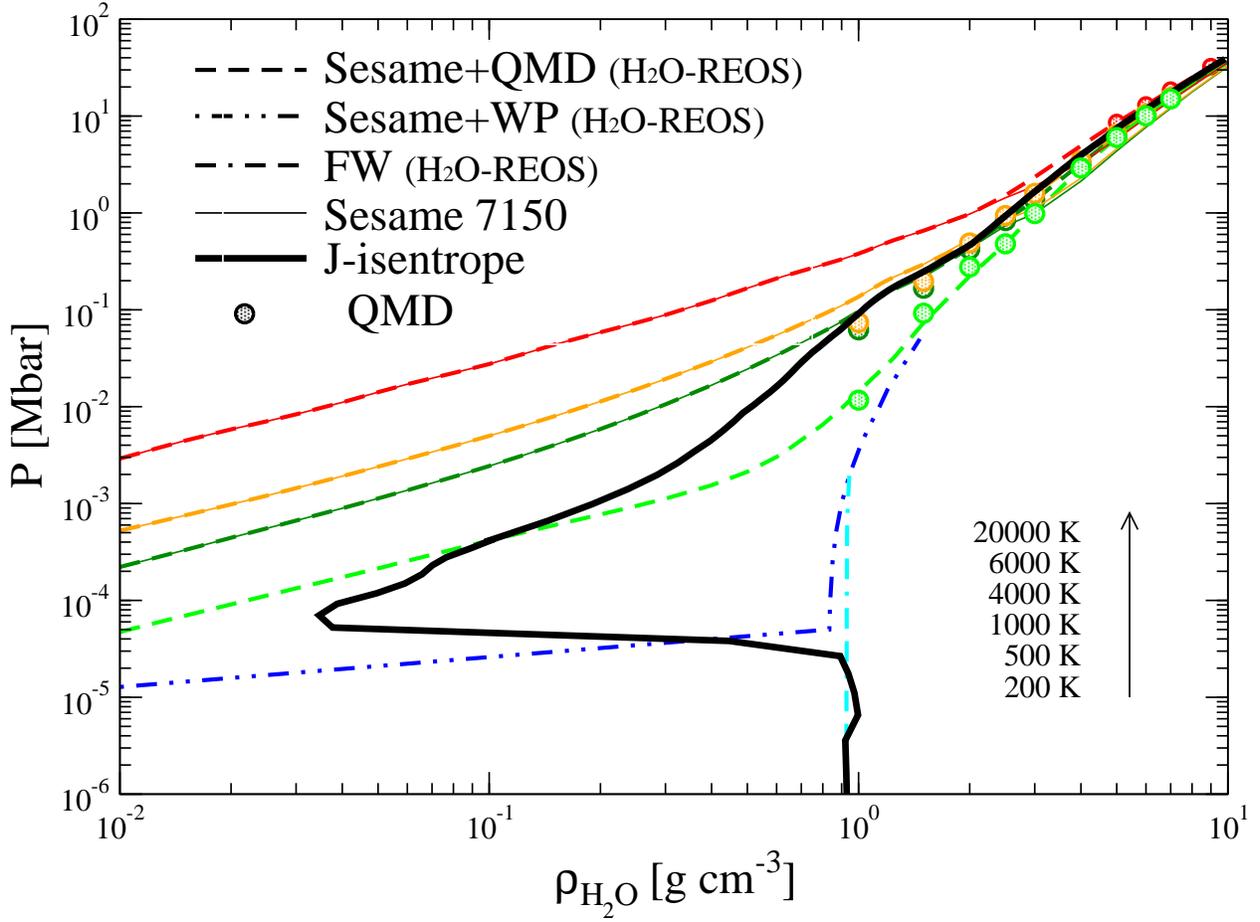}
%\plotone{./f3_nocolor.eps}
\figcaption{\label{fig_H2O_isoTs}(Color online) Isotherms for H$_2$O for T=200, 500, 1000, 4000, 
6000, and 20000~K according to different EOS tables: a combination of QMD and Sesame~7150 
for $T\geq 1000$~K, a combination of WP and Sesame~7150 for $T<1000$~K, FW \cite{FWeis} for ice~I, 
Sesame~7150, and QMD data. The thick line shows the pressure-density relation of the H$_2$O fraction 
along a typical Jupiter isentrope. All these isotherms are incorporated into H$_2$O-REOS.}
\end{figure}

Fig.~\ref{fig_H2O_isoTs} shows six pressure isotherms for H$_2$O-REOS 
representing different phases of H$_2$O and the pressure-density relation of the 
water fraction representing heavy elements along a typical Jupiter isentrope. Due 
to the phase diagram of pure H$_2$O, in the outermost region of Jupiter the 
H$_2$O-component transits from ice~I to liquid water below 300~K and further to 
the vapor phase around 550~K. At about 4000~K, H$_2$O again reaches densities of 
1~g\,cm$^{-3}$ along the Jupiter isentrope. Above this point, water shows strong 
molecular dissociation and becomes also electronically conductive~\citep{MD06}. 
From here on, the EOS of the water component in Jupiter is described  
by our QMD data. Following the isentrope further into the deep interior, water 
remains in the fluid, plasma-like phase. 

\placefigure{fig_H2O_isoTs}%f3.eps

Our H$_2$O phase diagram smoothly adds to recent QMD simulations by~\citet{MD06}. 
The EOS data differ at densities between 3-10 g~cm$^{-3}$ from the pure Sesame EOS.

%%%%%%%%%%%%%%%%%%%%%%%%%%%%%%%%%%%%%%%%%%%%%%%%%%%%%%%%%%%%%%%%%%
\subsection{EOS of metals and core material}\label{chap_EOSmetals}

\begin{figure}
\plotone{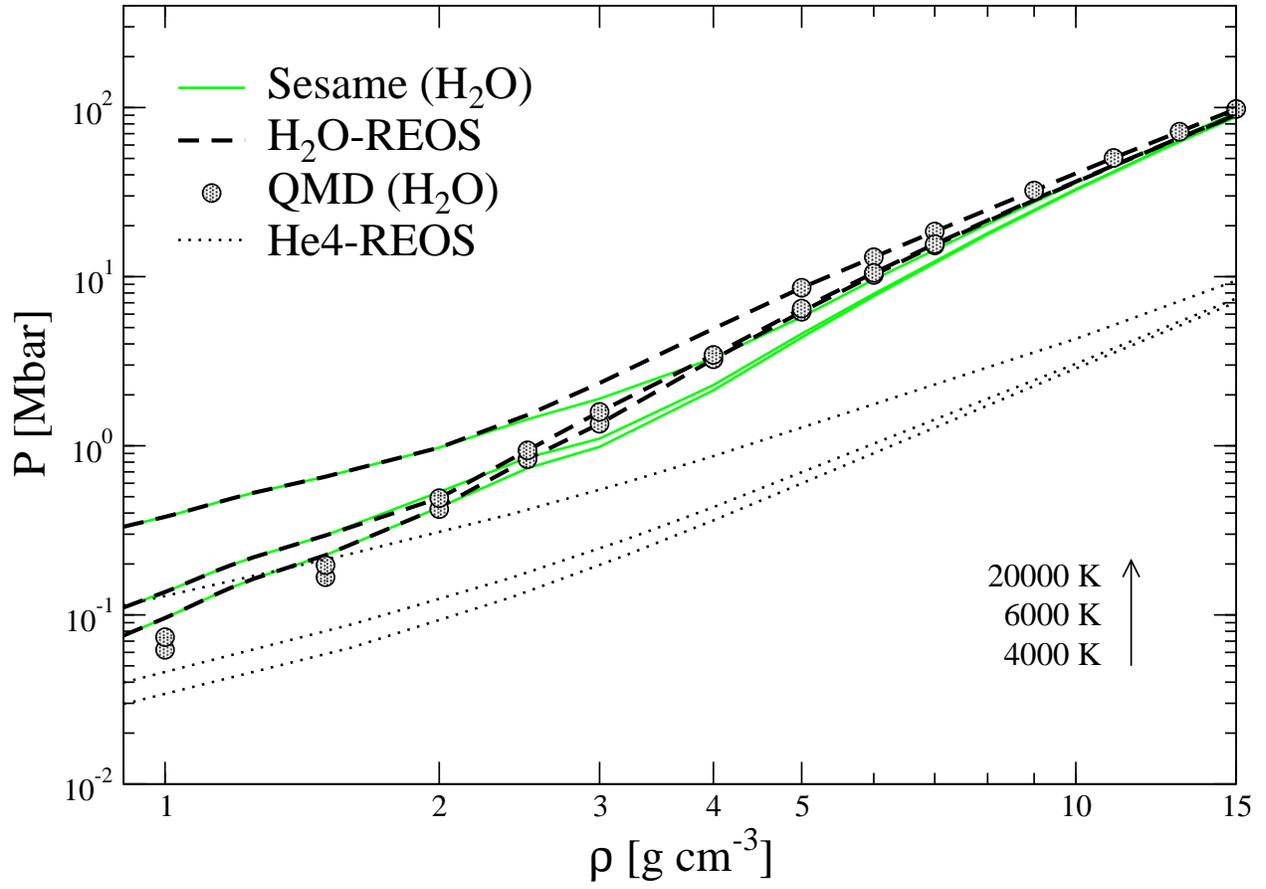}
%\plotone{./f4_nocolor.eps}
\figcaption{\label{fig_isoTs_Z}(Color online) Isotherms of the Z-component for T=4000, 6000, 
and 20000~K according to different EOS: Sesame~7150 for H$_2$O, QMD data for H$_2$O, 
H$_2$O-REOS, and He4-REOS.} 
\vspace*{0.2cm}
\end{figure}

We make use of the EOS table for water to represent metals in the envelopes. 
Alternatively, the EOS table of helium scaled in density by a factor of 4 (He4) is used 
to represent the atomic weight of oxygen. Oxygen is the third abundant element in the sun 
and probably also in Jupiter, see \citet{Ker04-1} for a discussion. We compare the respective 
isotherms of the EOS for metals in Fig.~\ref{fig_isoTs_Z} in that density and temperature region 
where QMD data of H$_2$O are applied. The H$_2$O pressures are up to 10~times higher than the 
He4 data, probably resulting from enhanced ideal and pronounced Coulomb contributions in water plasma. 

\placefigure{fig_isoTs_Z}%f4.eps

Previous models of Jupiter assumed that the core consists of rocks surrounded by 
ices~\citep{CSHL92,GZ99,Stevenson82}. Traditionally, rock material is a mixture of S, Si, Mg, 
and Fe while ice refers to a mixture of C, N, O, and H. Due to the core accretion scenario 
for the formation process of giant planets~\citep{Alibert04,Pollack96}, these elements have mainly
entered the planetary interior in form of icy compounds of CH$_4$, NH$_3$, 
H$_2$O, and of porous material composed of SiO$_2$, MgO, FeS, and FeO. Their present state 
in the deep interior of the planet cannot be inferred from observations or interior model 
calculations. Our QMD simulations show clearly that H$_2$O is in a plasma and not in a solid 
phase at conditions typical for the core of Jupiter~\citep{Fre+08}. Since CH$_4$ and NH$_3$ 
are more volatile than H$_2$O, they are expected to prefer the plasma phase too in the core 
region of Jupiter. Because a smaller-sized core results from interior models rather than from 
formation models of Jupiter, the core has been argued to erode with time~\citep{Alibert04,Gui03}. 
Motivated by the possibility of core erosion and by the miscibility of the icy component within 
the metallic envelope we assume that the core consists purely of rocks in all our calculations 
for Jupiter. The EOS of rocks is adapted from~\citet{HM89}.

%%%%%%%%%%%%%%%%%%%%%%%%%%%%%%%%%%%%%%%%%
\subsection{Mixtures}\label{chap_mixtEOS}

The EOS of a mixture with mass fractions $X$ of hydrogen, $Y$ of helium, and $Z$ of metals 
is calculated by means of the EOS for the pure components using the additive volume rule 
\citep{CSHL92,Peebles64} for the internal energy,
\begin{equation}
u(P,T) = Xu_X(P,T) + Yu_Y(P,T) + Zu_Z(P,T) \:,
\end{equation}
and the mass density,
\begin{equation}
\frac{1}{\rho(P,T)} = \frac{X}{\rho_X(P,T)} + \frac{Y}{\rho_Y(P,T)} + \frac{Z}{\rho_Z(P,T)} \:,
\end{equation}
on a grid of pressures $P$ and temperatures $T$. Non-ideality effects of mixing are thus 
not taken into account. 

In this paper we refer to three different linear mixtures of EOS data abbreviated by \textit{LM-REOS}, 
\textit{SCvH-ppt} and \textit{SCvH-i}. The new data set \textit{LM-REOS} consists of a linear 
mixture as described above of H-REOS, He-REOS, and of H$_2$O-REOS or He4-REOS for metals. 
\textit{SCvH-ppt} as used in this work consists of H-SCvH-ppt for hydrogen and He-SCvH for helium \citep{SCVH}, 
and of He-SCvH scaled in density by a factor four as representative for metals. \textit{SCvH-i} as used 
in this work consists of the interpolated version \citep{SCVH} of H-SCvH-ppt together with the same EOS 
for He and metals as for SCvH-ppt. Apart from SCvH-ppt, none of the hydrogen or helium EOS used here shows 
a first-order phase transition. 

For given mixtures, the piecewise isentrope defined by the boundary condition is calculated as described in 
\S\ref{chap_bequs}. When using H$_2$O for metals we make the following exception for temperatures below 1000 K. 
At these small temperatures where H$_2$O passes the phases ice I, liquid water, and vapor, we first calculate 
the isentrope of the H/He mixture and then add the desired mass fraction of H$_2$O to the $P-\rho$ relation 
of that isentrope according to its $P-T$ relation. At 1000 K, the pieces of isentropes for smaller and higher 
temperatures add very smoothly, as can be seen from Fig.~\ref{fig_H2O_isoTs}.

%%%%%%%%%%%%%%%%%%%%%%%%%%%%%%%%%%%%%%%%%%%%%%%%%%%%%%%%%%%%%%%%%%%%%
\subsection{Comparison with chemical models}\label{chap_cmpQMDvsSCvH}

Results of our LM-REOS for the coupling parameter $\Gamma$, the degeneracy parameter 
$\Theta$, and the pressure-density relation along the hydrogen adiabat are compared with the 
chemical models SCvH-ppt and SCvH-i.

\begin{figure}
\plotone{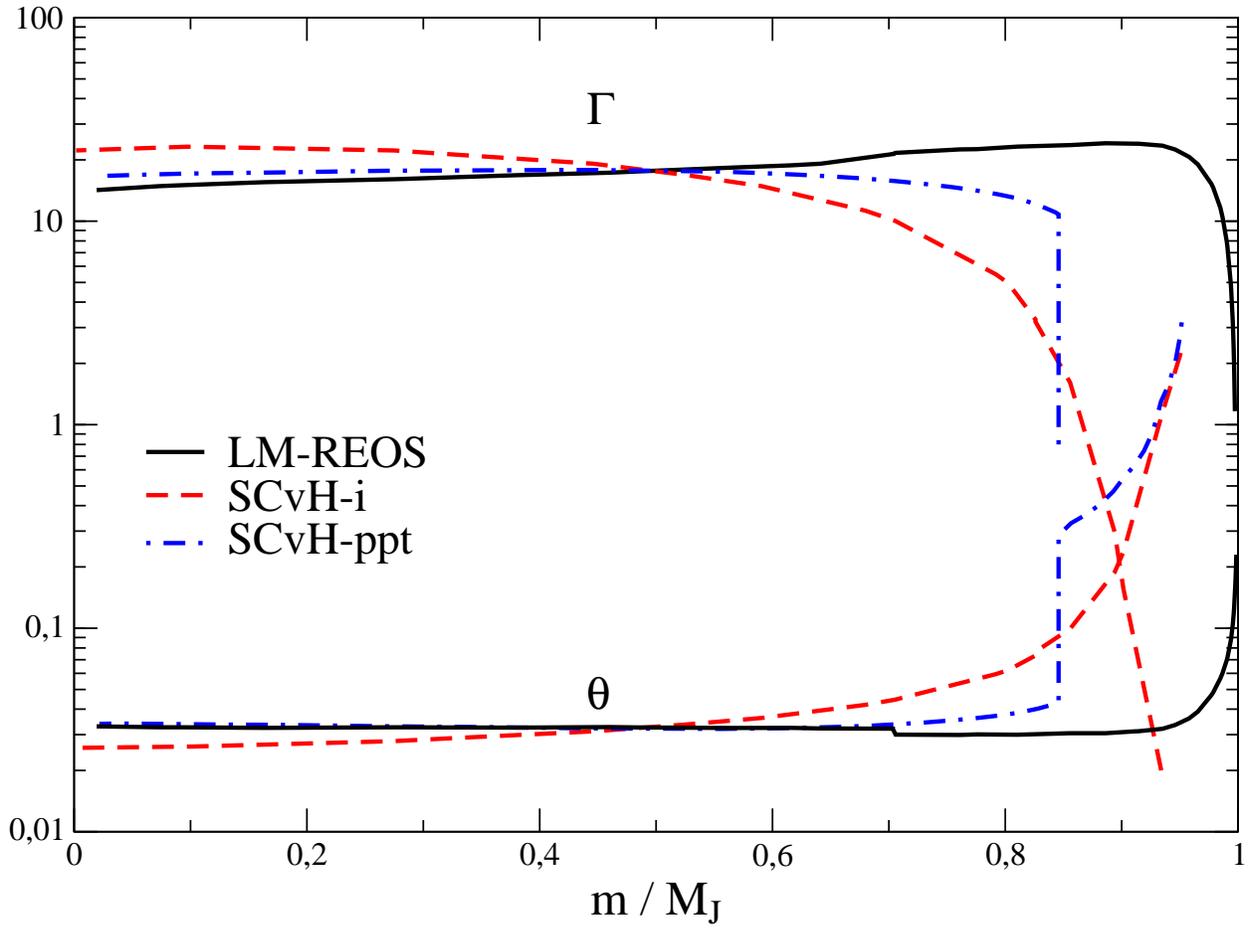}
%\plotone{./f5_nocolor.eps}
\figcaption{\label{fig_gammatheta}(Color online) Coupling parameter $\Gamma$ and degeneracy parameter $\Theta$ 
inside Jupiter for different EOS: LM-REOS, SCvH-i, and SCvH-ppt.}
\end{figure}

We have applied the definition of $\Gamma$ and $\Theta$ given by \citet{ChabPot} for multi-component plasmas 
with mean ion charge $\langle Z\rangle$ and electron density $n_e$
\begin{equation}
	\Gamma=\frac{e_0^2}{4\pi\epsilon_0k_B}\left(\frac{4\pi}{3}\right)^{1/3}
	\times \frac{\langle Z\rangle^{5/3}n_e^{1/3}}{T}
\end{equation}
\begin{equation}
	\Theta = \frac{k_B}{\hbar^2}2m_e(3\pi)^{2/3}\times\frac{T}{n_e^{2/3}}\:.
\end{equation}
These plasma parameters are calculated along typical Jupiter adiabats by taking into account 
the ionization degree of hydrogen and helium.
For models with the SCvH-ppt and SCvH-i EOS, the fraction of H$^+$, He$^+$, and He$^{2+}$ was 
taken simply from their tables. This is not directly possible for models using the QMD data sets 
because the underlying strict physical picture does not discriminate between bound and free states. 
However, we have estimated the ionization degree of the He component by using the COMPTRA04 program 
package~\citep{COMPTRA}. In the underlying chemical model of COMPTRA04, the composition of a partially 
ionized plasma at a given temperature and mass density is determined by the solution of coupled mass action 
laws including non-ideal contributions to the chemical potential via Pad{\'e}-formulas and a polarization 
potential accounting for the interaction between electrons and neutral atoms. For the hydrogen subsystem, 
the degree of dissociation is evaluated via the proton-proton distribution function and the coordination number 
derived from the QMD runs; this method is described in more detail by~\citet{Hol+08}. According to state-of-the-art 
chemical models for hydrogen such as SCvH or FVT$^+$, the fraction of atoms is always small so that we can estimate 
the ionization degree in the hydrogen component by taking the dissociation degree as an upper limit. Results for 
$\Gamma$ and $\Theta$ are shown in Fig.~\ref{fig_gammatheta} as function of the mass coordinate in Jupiter.

\placefigure{fig_gammatheta}%f5.eps

At least 80\% of Jupiter's mass is in a state with strong coupling and degeneracy effects, 
defined by $\Gamma>1$ and $\Theta<1$. With SCvH-ppt, the step-like increase of $\Gamma$ 
coincides with the density discontinuity at the layer boundary. LM-REOS has a much earlier 
onset of ionization in the hydrogen component than both SCvH-ppt and SCvH-i. In the very deep interior, 
only about 5\% of the He atoms are singly ionized and less than 1\% doubly ionized as inferred from the
chemical model COMPTRA04 applied to our LM-REOS data, while complete ionization is predicted by the 
SCvH-He EOS so that $\Gamma$ becomes slightly larger in that models. This large discrepancy between
chemical models in the region of pressure ionization is due to a different treatment of correlation effects
occuring in the exponents of the mass-action laws for ionization, from which the composition is derived. 
Higher temperatures as with the SCvH-ppt EOS diminish the degeneracy and in spite of fully ionized helium 
the same values are found as with LM-REOS.

\begin{figure}
\plotone{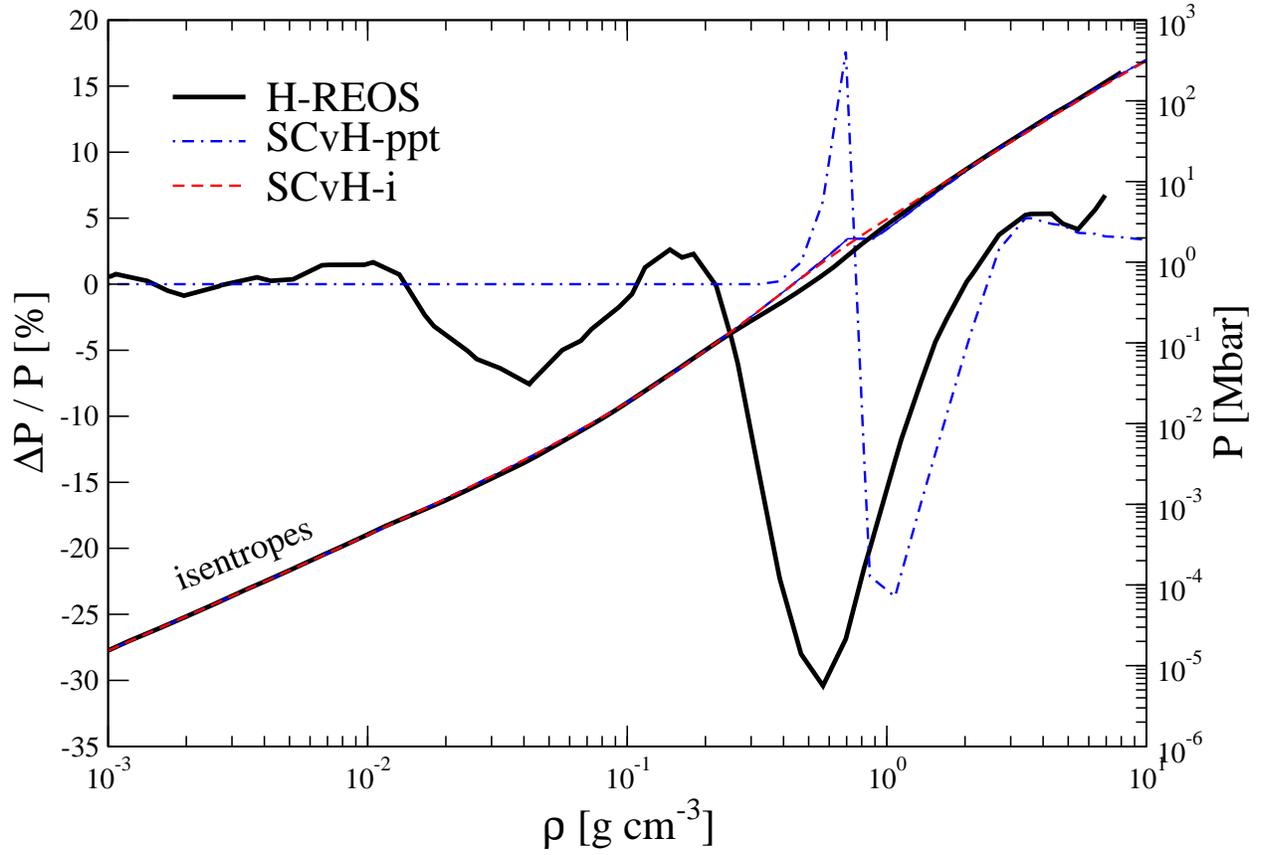}
%\plotone{./f6_nocolor.eps}
\figcaption{\label{fig_isen}(Color online) Hydrogen adiabats for Jupiter, determined by T=165~K at P=1~bar, computed 
with three different hydrogen EOS: H-REOS, SCvH-i, and SCvH-ppt. The scale on the right shows absolute pressures and 
the scale on the left shows relative differences in pressure with respect to the SCvH-i-adiabat.}
\end{figure}

%%%% Isentorpen
More than 50\% of Jupiter's mass is in a high-pressure state between 0.1 and 10~Mbar where 
the differences in the hydrogen isentropes are most pronounced and manifest themselves in 
the structure models, see \S\ref{chap_resJupQMD}. With the transition from FVT$^+$ data to 
QMD data around 0.05~Mbar, the Jupiter isentrope of pure hydrogen becomes much softer than H-SCvH-i, 
with a maximum difference of 30\% at 1~Mbar, as shown in Fig.~\ref{fig_isen}. 
In the innermost 20\% of Jupiter's mass, i.e.\ above 20~Mbar, the H-REOS isentrope follows 
the H-SCvH-ppt one with 5\% higher pressures than SCvH-i. Differences below 0.1~Mbar do 
not significantly contribute to the structure models since this region occupies less than 
2\% of Jupiter's total mass.

\placefigure{fig_isen}%f6.eps

%%%%%%%%%%%%%%%%%%%%%%%%%%%%%%%%%%%%%%%%%%%%%%%
\section{Results for Jupiter}\label{chap_resJup}

In this section we present results for Jupiter's core mass, for the abundances of metals 
in the envelopes, and for the profiles of the main components along the radius. The size 
of the core $M_c$ determined by the condition of total mass conservation, and the abundances 
of metals $Z_{\rm mol}$, $Z_{\rm met}$ which are needed to reproduce $J_2$ and $J_4$ are very 
sensitive with respect to the choice of the EOS and the precise values of the observational 
parameters to be reproduced. First, we compare our results with those of~\citet{Gui03} 
by using the same EOS data sets (SCvH-ppt and SCvH-i) in order to demonstrate that our 
procedure is sound. In \S\ref{chap_resJupQMD} we then present the range of acceptable Jupiter 
models using our new LM-REOS in comparison with SCvH-based models, explore in \S\ref{chap_resJupSystematic} 
the behavior of LM-REOS based models by varying $T_1$, $J_4$, and $P_{\rm m}$, and discuss in \S\ref{chap_torten} 
the fractions of chemical species calculated with LM-REOS and SCvH-ppt for Jupiter models which satisfy the same 
observational constraints given in Tab.~\ref{tab_obs}.

%%%%%%%%%%%%%%%%%%%%%%%%%%%%%%%%%%%%%%%%%%%%%%%%%%%%%%%%%%%%%%%%%%%%%%
\subsection{Consistency with former results}\label{chap_resJupTable}

We follow the notation of~\citet{Gui03} who has performed extensive calculations 
for planetary interiors and study four of the models he has introduced explicitly. These 
reference models (A, B, D, E)-TG are based on a three-layer structure as also adapted here. 
Tab.~\ref{tab_ABDE} contains their first independent confirmation. 
In our recalculated models labeled (A,\ldots, E)-this, we do not allow for 
a deviation of $\bar{Y}$ and $J_4$ of more than 1/10$\sigma$; but in our models labeled (A',\ldots, E')-this, $\bar{Y}$ and $J_4$ are varied within 1$\sigma$ in order to achieve the best agreement as possible. Corresponding numbers are listed in Tab.~\ref{tab_ABDEparam}. 

\placetable{tab_ABDE}
\clearpage
\begin{deluxetable}{lcccccc}
\tablecaption{\label{tab_ABDE}Comparison of present (this) and former (TG) results.}
\tablehead{
	\colhead{Model} & \colhead{H-EOS} & \colhead{$T_1$} & \colhead{$M_c$} & \colhead{$M_Z$} & 
	\colhead{$Z_{\rm met}$} & \colhead{$Z_{\rm mol}$}\\	
	\colhead{} & \colhead{SCvH-} & \colhead{[K]} & \colhead{$[M_{\oplus}]$} & \colhead{$[M_{\oplus}]$} & 
	\colhead{$[Z_{\odot}]$} & \colhead{$[Z_{\odot}]$}	
}
\startdata
A-TG			&i		&165			&4.2			&33.1		&4.7					&5.3		\\
A-this		&i		&165			&0.7			&35.6		&6.0					&4.5		\\	
A'-this		&i		&165			&2.8			&34.2		&5.2					&5.0		\\
\tableline
B-TG			&i		&170			&0				&35.3		&6.1					&4.1		\\
B-this		&i		&170 		&0.8			&36.4		&6.0					&4.9		\\	
B'-this		&i		&170			&0				&35.5		&6.0					&4.8		\\
\tableline
E-TG			&ppt	&165			&4.3			&19.4		&2.5					&2.3		\\
E-this		&ppt	&165 		&6.7			&24.8		&2.9					&3.9		\\
E'-this		&ppt	&165			&4.6			&22.7		&2.9					&3.3		\\ 
\tableline
D-TG			&ppt	&170			&10.0		&17.5		&0.7					&4.5		\\
D-this		&ppt	&170 		&7.0			&25.7		&2.9					&4.2		\\
D'-this		&ppt	&170			&8.6			&22.6		&1.9					&4.9		\\
\enddata
\tablecomments{The mass fraction of heavy elements is given in solar units $Z_{\odot}=1.92\%$. Models (A--E)-TG are taken from \citet{Gui03}. Present models with prime are designed to give the best agreement, see Table~\ref{tab_ABDEparam} for the choice of parameters.}
\end{deluxetable}

\clearpage

Models (A--E)-TG indicate that $T_1$ affects $M_c$ by several $M_{\oplus}$, but 
apparently not systematically. Keeping other constraints constant, we always find a slightly 
higher core mass for a larger 1-bar temperature. On the other hand, shifting the layer boundary 
between the envelopes outwards, e.g.\ from 2.0 to 1.4~Mbar in model A'-this, always magnifies the core mass such 
that the result $M_c(A)>M_c(B)$ of \citet{Gui03} can be reproduced. Our value of $M_c=2.8$~$M_{\oplus}$ 
in model A'-this can be further increased by 0.6~$M_{\oplus}$ if we use ices as core material instead of 
rocks.

\placetable{tab_ABDEparam}
\clearpage
\begin{deluxetable}{ccccccc}
\tablecaption{\label{tab_ABDEparam}Parameters of present Jupiter models using SCvH EOS}
\tablehead{
\colhead{Model} & \colhead{$P_{\rm m}$	[Mbar]} & \colhead{$\bar{Y}$ [\%]} & \colhead{$J_4/10^{-4}$}}	
\startdata
A\&B-this	& 2.00	& 27.5	& -5.84\\
E-this	& 1.73	& 27.5	&-5.84\\
D-this	& 1.76	& 27.5	&-5.84\\
\tableline
A'-this	& 1.40	& 28.0	& -5.89\\
B'-this	& 2.00	& 27.8	& -5.82\\
E'-this	& 1.77	& 28.5	& -5.79\\
D'-this	& 1.74	& 28.5	& -5.89\\
\enddata
\tablecomments{In models using SCvH-ppt EOS, $P_{\rm m}$ coincides with the PPT. These models have been calculated in order to check the consistency with former results, see Table~\ref{tab_ABDE}.}
\end{deluxetable}

\clearpage

In agreement with \citet{Gui03} the total abundance of metals $M_Z=M_{Z_{\rm mol}}+M_{Z_{\rm met}}+M_c$ increases by 10-15~$M_{\oplus}$ when using SCvH-i instead of SCvH-ppt. Since all our solutions yield $Z_{\rm mol}<Z_{\rm met}$ with SCvH-i and $Z_{\rm mol}>Z_{\rm met}$ with SCvH-ppt in agreement with the bulk of solutions in \citet{Gui03}, the inverse ratio of models A-TG and E-TG cannot be reproduced. Apart from the deviations of 43\% for $Z_{\rm mol}$(E') and 170\% for $Z_{\rm met}$(D'), the deviations of all other solutions are below 30\% so that they lie within the range of solutions found in~\citet{Gui03}. 
With $0<M_c<10$, $17.5<M_Z<35.3$, $2.3<Z_{\rm mol}<5.3$, and $0.7<Z_{\rm met}<6.1$ the solutions 
(A--E)-TG span large ranges in units of Earth masses and solar abundances $Z_{\odot}=1.92\%$, 
respectively. Our nearest results differ by up to 1.4~$M_{\oplus}$ for $M_c$, 5.1~$M_{\oplus}$ 
for $M_Z$, 1.0~$Z_{\odot}$ for $Z_{\rm mol}$, and 1.2~$Z_{\odot}$ for $Z_{\rm met}$ from these results. To reduce the disagreement in model $D$ with respect to $Z_{\rm met}$ and $M_Z$, which are nearer to model E-TG than to D-TG, the mean helium content should be higher or the EOS of metals be more compressible. 

We conclude that our method is able to reproduce former results well enough to smoothly continue and complement investigations of the interplay between the internal structure of giant planets and the EOS applied.

%%%%%%%%%%%%%%%%%%%%%%%%%%%%%%%%%%%%%%%%%%%%%%%%%%%%%%%
\subsection{New results with LM-REOS}\label{chap_resJupQMD}

\begin{figure}
\plotone{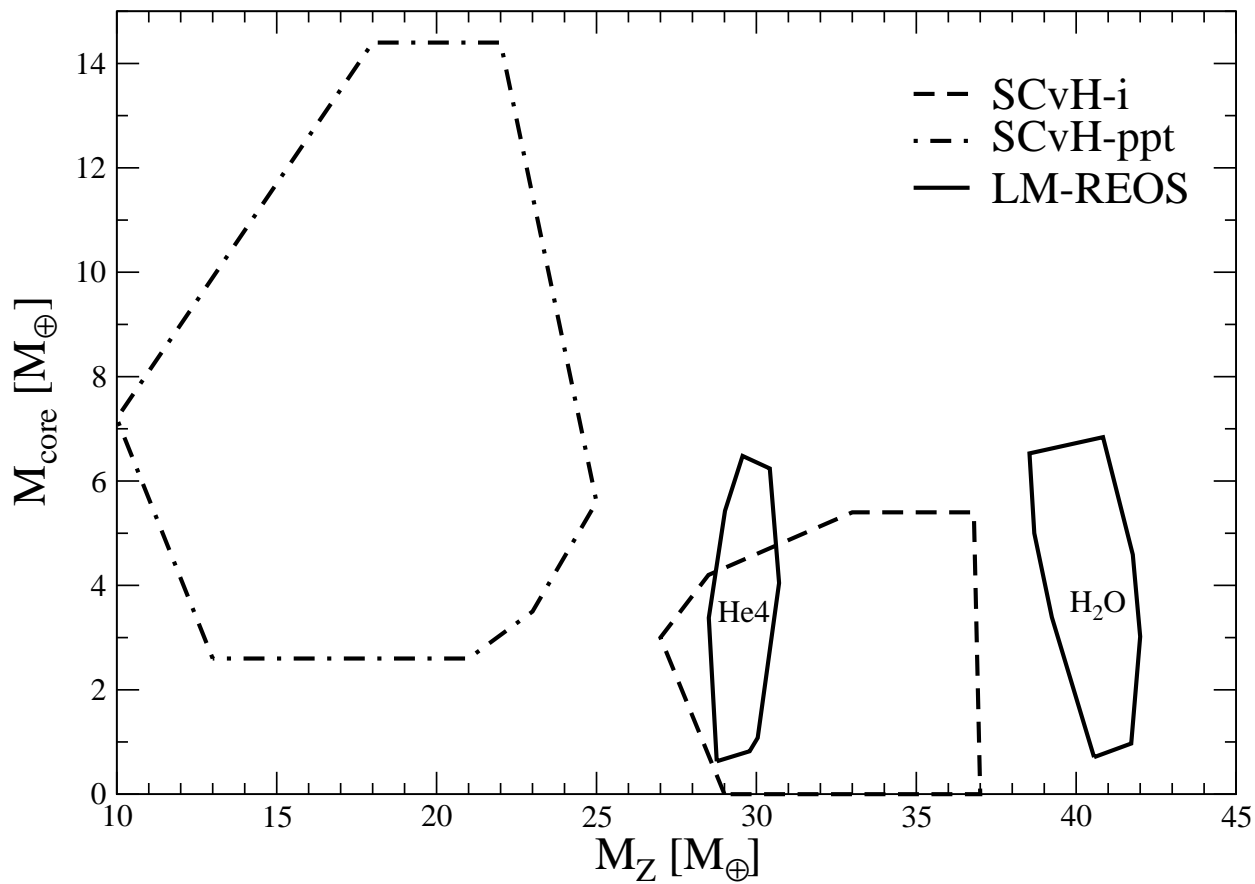}
\figcaption{\label{fig_McZ}
Core mass and total mass $M_Z$ of heavy elements in units of Earth masses for three different EOS. Solutions 
with LM-REOS (solid lines) demonstrate a strong influence of the choice of the EOS of metals to the resulting total 
heavy element abundance and are displayed in seperate boxes labeled 'He4' and 'H$_2$O', respectively.
}
\end{figure}

\begin{figure}
\plotone{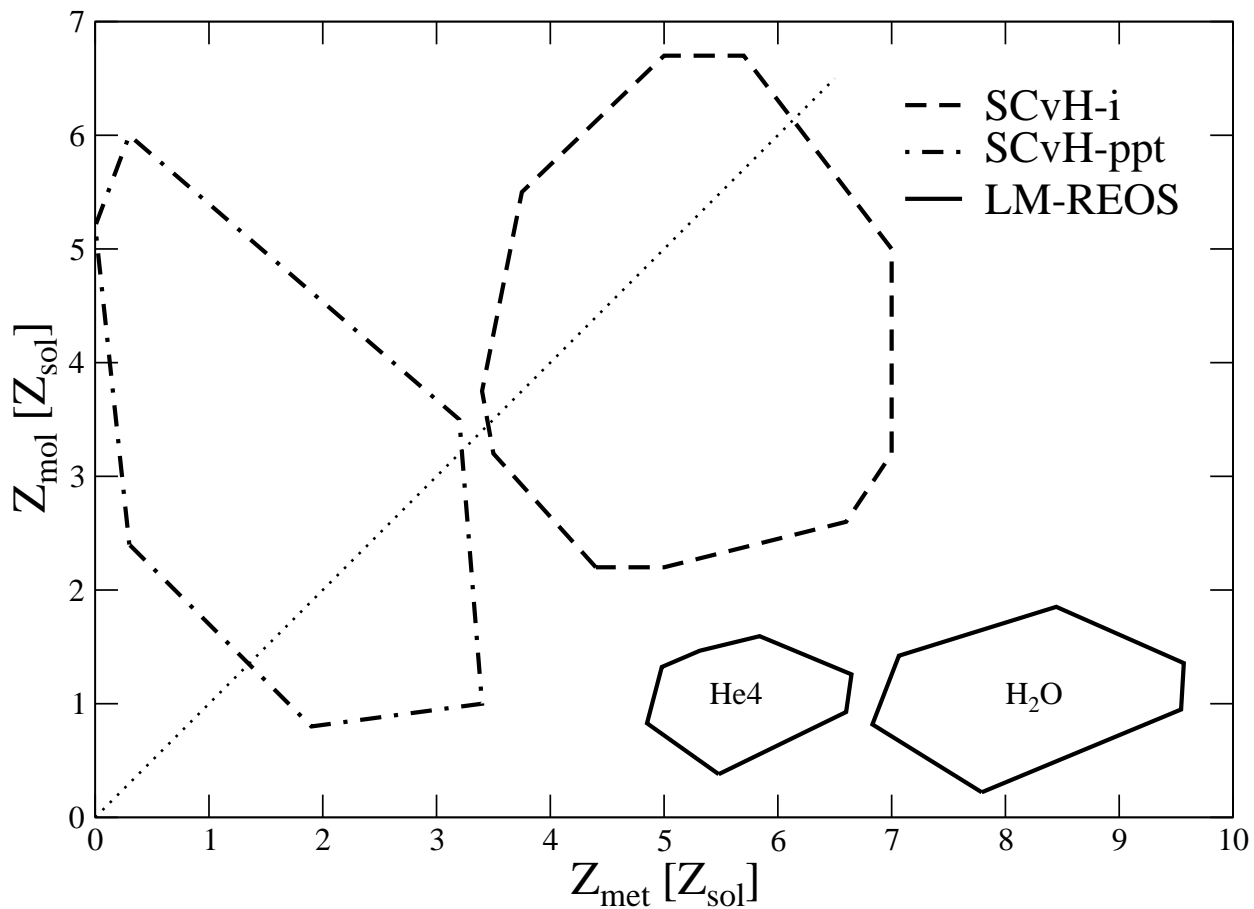}
\figcaption{\label{fig_ZZ}
Abundance of heavy elements in solar units in the molecular layer ($Z_{\rm mol}$) and in the metallic layer 
($Z_{\rm met}$) for three different EOS. The dotted line indicates equal abundances. Solutions with LM-REOS 
(solid lines) demonstrate a strong influence of the choice of the EOS of metals to the resulting value of
$Z_{\rm met}$ and are displayed in seperate boxes labeled 'He4' and 'H$_2$O', respectively.}
\end{figure}

Figures~\ref{fig_McZ} and \ref{fig_ZZ} show the full range of solutions using QMD data in comparison with models based on SCvH-ppt and SCvH-i adapted from~\citet{Gui03} with respect to $M_c$, $M_Z$, $Z_{\rm mol}$, and $Z_{\rm met}$. Figures~\ref{fig_McZ_QMD} and \ref{fig_ZZ_QMD} illustrate the effect of a varying $P_{\rm m}$, $T_1$, and $J_4$ on the position of the QMD data based solutions. For all our solutions, $Y_{\rm mol}=0.238$, $\bar{Y}=0.275$, $P_{\rm m}$ is varied from 3 to 5~Mbar, $T_1$ from 165 to 170~K and $J_4$ from -5.84 to -5.89$\times 10^{-4}$. Metals are represented either by He4 or by H$_2$O, see \S\ref{chap_EOSmetals}.

The behavior of the isentropes already indicates (see Fig.\ \ref{fig_isen}) that the solutions 
based on LM-REOS are nearer to the SCvH-i than to the SCvH-ppt results. This is a clear 
consequence of the missing PPT in the H-REOS data. The PPT makes the H-EOS less compressible 
in the molecular regime and more compressible in the metallic regime before it is dominated by 
electronic degeneracy. 

The core mass of QMD-based models does not vanish for transition pressures up to 5~Mbar and 
$J_4 \leq J_4^{\rm obs}$. On the other hand, the core mass of our SCvH-i models does not 
exceed 4~$M_{\oplus}$ at transition pressures as low as 1~Mbar. Higher core masses could be 
achieved with SCvH-i for $P_{\rm m}<1$~Mbar, but are not reasonable since the layer boundary is to  represent the PPT which occurs above 1~Mbar for SCvH-ppt. Higher core masses of QMD-based 
models can be explained by a smaller compressibility at pressures above 10~Mbar as seen from the 
isentropes. For a given abundance of metals, a smaller compressibility of hydrogen adds less 
mass to the isentrope and leaves more mass for the core. Similary, the smaller the 
compressibility of the EOS of metals (Fig.\ \ref{fig_isoTs_Z}), the higher their abundance 
needed to contribute mass to the isentrope in order to match $J_2$ and $J_4$.

\placefigure{fig_McZ}%f7.eps

A pronounced difference between the isentropes calculated with the LM-REOS and SCvH-i occurs 
around 1~Mbar, where the QMD pressure is up to 30\% smaller. This feature leads to smaller 
$Z_{\rm mol}$ and higher transition pressures, see Fig.\ \ref{fig_ZZ}, which can be explained as 
follows. For a given pressure-density relation along the isentrope as required by 
equations~(\ref{eq_dPdl}) and (\ref{eq_J2n}), a higher compressibility of the hydrogen component 
has to be compensated by a smaller metal abundance in order to keep $J_2$ and $J_4$ at the 
observed values. Guillot~\citeyearpar{Guillot05} has also shown that the mean radius of the contribution 
of level radii to the gravitational moments $J_{2n}$ increases with increasing index $n$ and is 
placed around the layer boundary in case of $J_2$ and $J_4$, whereas the contribution to higher 
moments than $J_2$ from the deep interior is negligible. The molecular region contributes slightly more 
to $J_4$ than to $J_2$, and the metallic layer contributes slightly more to $J_2$ than to $J_4$. 
If the compressibility of hydrogen becomes large near the molecular-metallic layer boundary, 
$Z_{\rm mol}$ has to be chosen small in order to not exceed $|J_4|$; but simultaneously $J_2$ will 
become too low and thus $Z_{\rm met}$ has to be enhanced strongly, again enlarging simultaneously 
$|J_4|$ beyond the desired value. Therefore, in order to adjust both gravitational moments with a 
$Z_{\rm mol}$ not too small, the influence of $Z_{\rm met}$ on $J_4$ has to be reduced by shifting 
the layer boundary inwards. 
LM-REOS based models of Jupiter have $Z_{\rm mol}=0.2-1.8\,Z_{\odot}$, $Z_{\rm met}=4.8-9.6\,Z_{\odot}$ 
for $P_{\rm m}=$ 3 to 5~Mbar. 

\placefigure{fig_ZZ}%f8.eps

%%%%%%%%%%%%%%%%%%%%%%%%%%%%%%%%%%%%%%%%%%%%%%%%%%%%%%%
\subsection{Systematic behavior of LM-REOS solutions}\label{chap_resJupSystematic}

\begin{figure}
\plotone{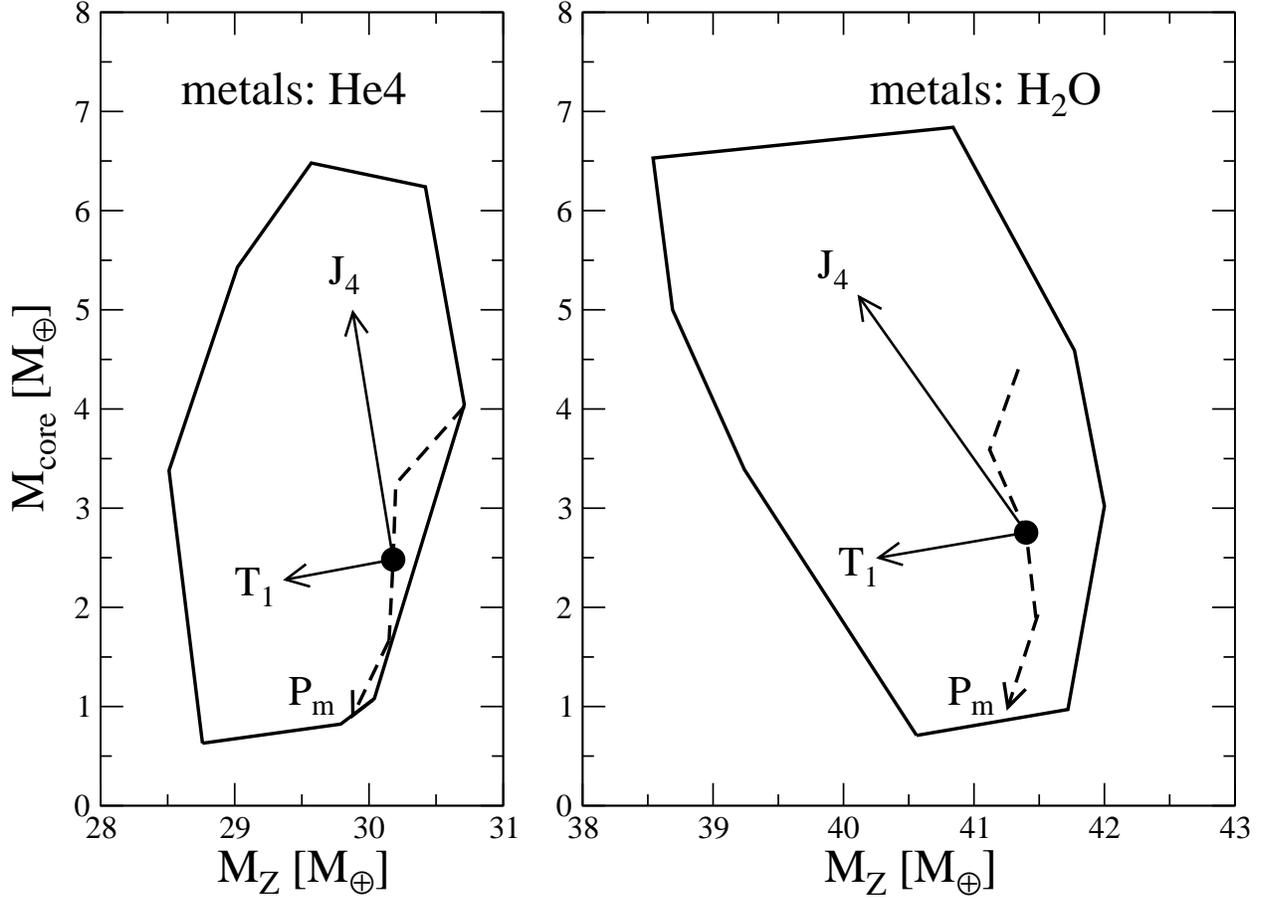}
\figcaption{\label{fig_McZ_QMD}
Same as Fig.\ \ref{fig_McZ} but only for solutions using LM-REOS. The arrows indicate the shifts of the reference 
solution ($T_1=170$~K, $\bar{Y}=0.275$, $P_{\rm m}=4$~Mbar, $J_4/10^4=-5.84$) if $T_1$ is decreased to 165~K, 
$|J_4|$ increased by 1~$\sigma$, or $P_{\rm m}$ enhanced from 3 to 5~Mbar.}
\end{figure}

\begin{figure}
\plotone{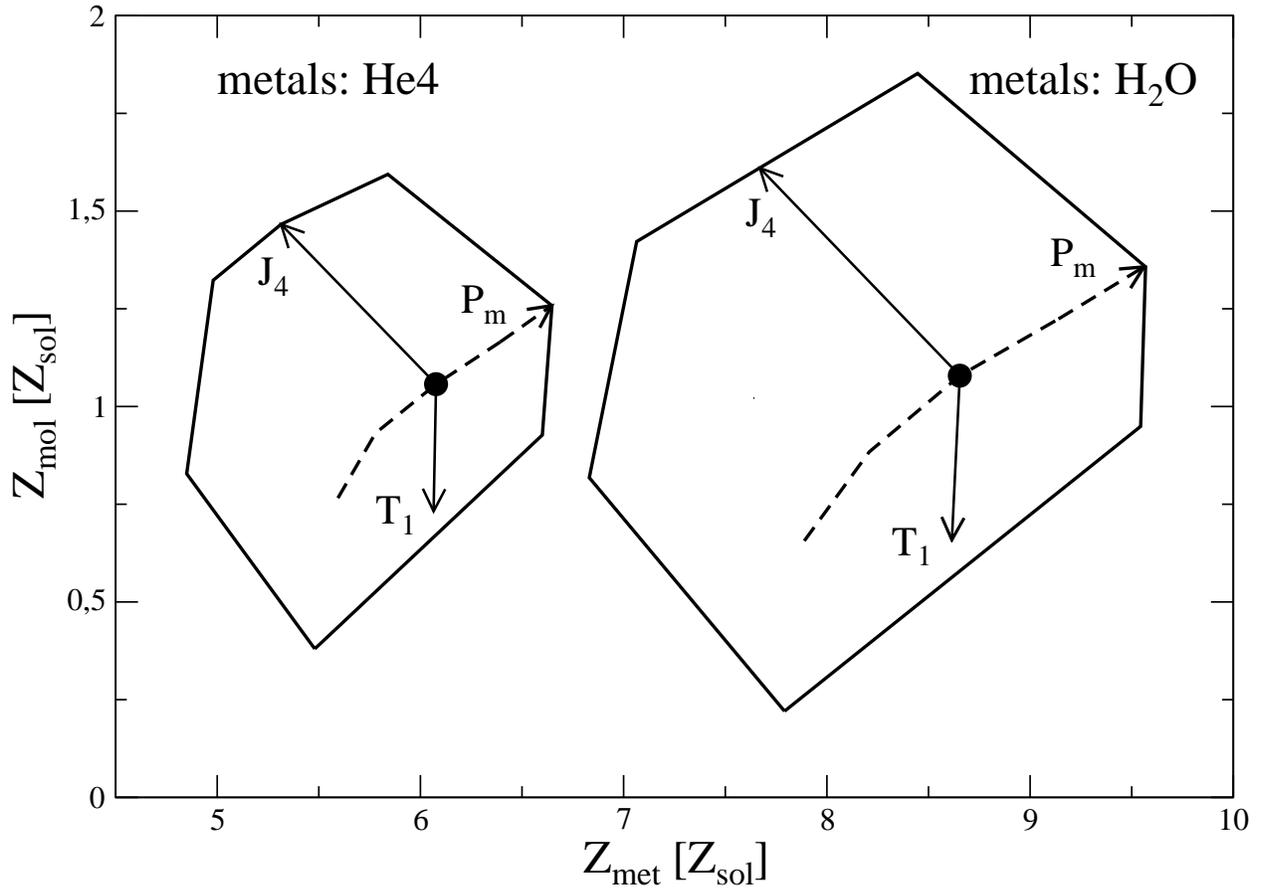}
\figcaption{\label{fig_ZZ_QMD} 
Same as Fig.\ \ref{fig_ZZ} but only for solutions using LM-REOS. See Fig.\ \ref{fig_McZ_QMD} for a description.}
\end{figure}

In this paragraph we illustrate the influence of the uncertainties in observational and free 
parameters on the position of the solution of LM-REOS based models in the usual $M_c$ vs.\ $M_Z$ 
and $Z_{\rm mol}$ vs.\ $Z_{\rm met}$ diagrams. For our point of reference, displayed as filled circle 
in Figs.\ \ref{fig_McZ_QMD} and \ref{fig_ZZ_QMD}, we have chosen the values $J_4/10^{-4}=-5.84$, 
$T_1=170$~K, and $P_{\rm m}=4$~Mbar. As explained above, $Z_{\rm mol}$ increases with $|J_4|$. 
We do not consider solutions for $|J_4|/10^4<5.84$ since they result in $Z_{\rm mol}<Z_{\odot}$ or 
$M_c=0$. In the molecular envelope where the isentrope is more sensitive with respect to temperature 
than in the degenerate metallic region, a cooler interior initiated by a smaller $T_1$ enhances 
the partial density of the He-H mixture and reduces the fraction of metals by 1~$M_{\oplus}$. 
Most important, the solutions are affected by the choice of the transition pressure. For 
$P_{\rm m}<3$~Mbar, no solution exists with $Z_{\rm mol}>Z_{\odot}$. Solutions with $1<Z_{\rm mol}<1.8\times Z_{\odot}$ 
are found for $3<P_{\rm m}<5$~Mbar in combination with $T_1=170$ K and $J_4\leq-5.84\times 10^{-4}$. 
A slightly higher enrichment factor of 2.0 is possible for $P_{\rm m}=7$~Mbar, before the core mass 
shrinks to zero. In contrast to the large influence of $P_{\rm m}$ on $Z_{\rm mol}$ and $Z_{\rm met}$, 
$M_Z$ just varies within one~$M_{\oplus}$. Finally, representing metals by H$_2$O instead of He4 
enhances $Z_{\rm met}$ by 50\% and shifts $M_Z$ from about 30 to about 40~$M_{\oplus}$. 

\placefigure{fig_McZ_QMD}%f9.eps
\placefigure{fig_ZZ_QMD}%f10.eps

%%%%%%%%%%%%%%%%%%%%%%%%%%%%%%%%%%%%%%%%%%%%%%%%%%%%%%%%%%%%%%%
\subsection{Profiles of chemical components}\label{chap_torten}

\begin{figure}
\plotone{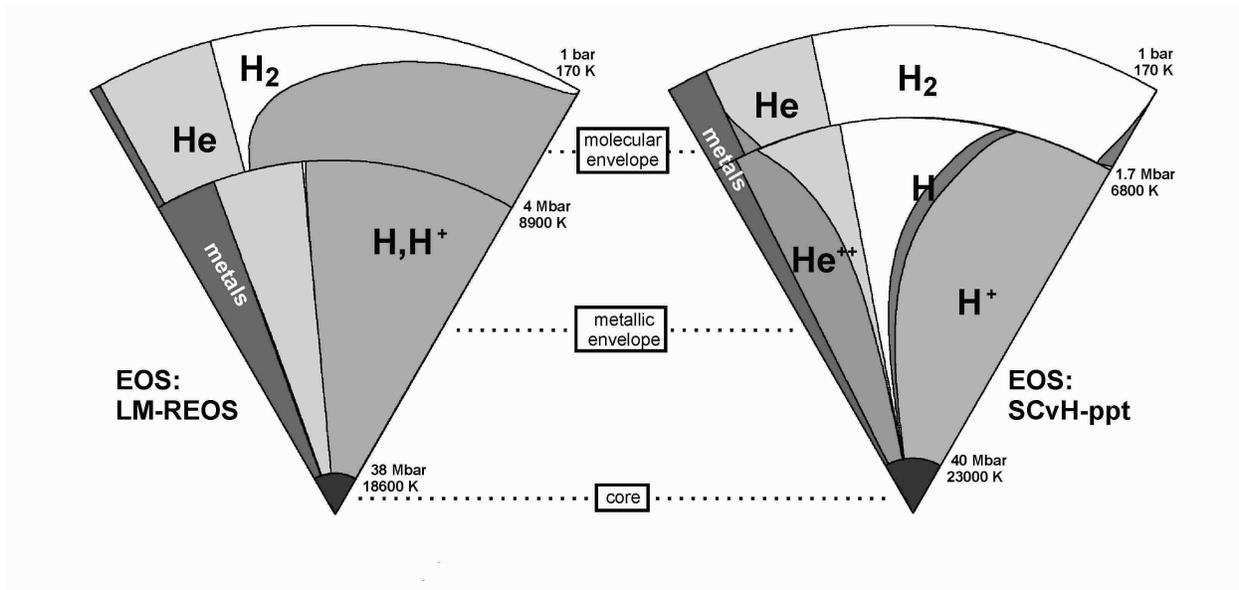}
\figcaption{\label{fig_torten}  
Schemes of Jupiter models satisfying the same constraints; left: model J\ref{fig_torten}a, right: model 
J\ref{fig_torten}b. At the layer boundaries the values of pressure and temperature are given. The abundances 
of metals and of chemical species along the radius are indicated by grey scales. An arc segment corresponds to 
100\% in mass. For model J\ref{fig_torten}a, an ASCII data table containing the profiles of pressure, temperature, 
density, composition and the figure functions along the radius can be found in the electronic edition of this Journal.} 
\end{figure}

We compare the abundances of single chemical species, see \S\ref{chap_cmpQMDvsSCvH}, along the radius 
inside Jupiter as calculated with LM-REOS (model J\ref{fig_torten}a in Fig.~\ref{fig_torten}, 
corresponding to the point of reference for H$_2$O in Fig.~\ref{fig_McZ_QMD}) and with SCvH-ppt 
(model J\ref{fig_torten}b in Fig.~\ref{fig_torten}). In these figures, a full arc segment corresponds 
to 100\% in mass. The radius coordinate is not displayed here but scales linearly from the center 
to the pressure level of 1~bar. In both of the models the core consists of rocks, $J_4/10^{-4}=-5.84$, 
$\bar{Y}=27.5$\%, and $T_1=170$~K. There is no degree of freedom in model J\ref{fig_torten}b and the 
only degree of freedom in model J\ref{fig_torten}a, $P_{\rm m}$, is set to 4 Mbar in order to give 
$Z_{\rm mol}>1$. 
The three most important differences are: (i) the steep onset of dissociation and ionization of H-atoms 
with the PPT in model J\ref{fig_torten}b wheras LM-REOS yields a smooth increase of dissociation; 
(ii) a high transition pressure of 4~Mbar in model J\ref{fig_torten}a, well beyond the location of the 
PPT in SCvH-ppt; (iii) a ratio of heavy elements $Z_{\rm mol}$:$Z_{\rm met}$ of 4:3 in model 
J\ref{fig_torten}b compared to 1:6 with LM-REOS. It is interesting to note that once neutral H and 
He$^+$ are formed in the SCvH-ppt model, they are almost immediately ionized further, whereas the 
fraction of ionized He of only 6\% in the deep interior in model J\ref{fig_torten}a is not resolved. 

\placefigure{fig_torteQMD}%f11.eps

For model J\ref{fig_torten}a, an ASCII data table containing the profiles of pressure, temperature, 
density, composition and the figure functions along the radius is availiable in the electronic edition 
of this Journal. A shortened version of this data table is shown in Tab. \ref{tab_J11a}, where the 
five rows, from top to bottom, present the 1-bar-level surface, the transition from the outer to the 
inner envelope, and the layer transition from the inner envelope to the core. 

\clearpage

\begin{deluxetable}{rrrrrrrrrrrrr}
\tabletypesize{\scriptsize}
\rotate
\tablewidth{0pt}
\tablecaption{\label{tab_J11a}Jupiter model J11a}
\tablehead{
\colhead{$m$} & \colhead{$P$} & \colhead{$l$} & \colhead{$T$} & \colhead{$\rho$} & 
\colhead{Y} & \colhead{Z} & \colhead{X$_{\rm H_2}$} & \colhead{X$_{\rm He}$} & \colhead{X$_{\rm He^+}$} & 
\colhead{$s_2$} & \colhead{$s_4$} & \colhead{$s_6$}\\
\colhead{[M$_{\oplus}$]} & \colhead{[Mbar]} & \colhead{[$\bar{R}_{\rm J}$]} & \colhead{[K]} & 
\colhead{[g\,cm$^{-3}$]} & \colhead{[\%]} & \colhead{[\%]} & 
\colhead{} & \colhead{} & \colhead{} & \colhead{} & \colhead{} & \colhead{}
}	
\startdata
317.8336 & 1.0000E-06 & 1.00000 & 1.7000E+02 & 1.6653E-04 & 23.307 & 2.072 & 1.0000 & 1.0000 & 0.0000 & -4.501E-02 & 1.984E-03 & -2.470E-04\\
\vdots & \vdots & \vdots & \vdots & \vdots & \vdots & \vdots & \vdots & \vdots & \vdots & \vdots & \vdots & \vdots\\
223.6070 & 3.9998E+00 & 0.72384 & 8.8682E+03 & 1.3239E+00 & 23.307 & 2.072 & 0.0095 & 1.0000 & 0.0000 &  -3.522E-02 & 1.062E-03 & -1.742E-04\\
223.5810 & 4.0013E+00 & 0.72380 & 8.8694E+03 & 1.5253E+00 & 24.476 & 16.616 & 0.0095 & 1.0000 & 0.0000 &  -3.522E-02 & 1.062E-03 & -1.744E-04\\
\vdots & \vdots & \vdots & \vdots & \vdots & \vdots & \vdots & \vdots & \vdots & \vdots & \vdots & \vdots & \vdots\\
2.75441 & 3.8385E+01 & 0.08426 & 1.8571E+04 & 4.3293E+00 & 24.476 & 16.616 & 0.0000 & 0.9405 & 0.0574 & -1.160E-02 & 1.063E-04 & -7.278E-06\\
2.75308 & 3.8393E+01 & 0.08426 & 1.8572E+04 & 1.8037E+01 & 0.000 & 0.000 & 0.0000 & 0.0000 & 0.0000 & -1.160E-02 & 1.063E-04 & -7.278E-06 
\enddata
\tablecomments{This table is a truncated version of a machine-readable table that is published in its 
entirety in the electronic edition of the Astrophysical Journal.
Column headings from left to right: mass coordinate, pressure, level coordinate (radius, see Eq.~\ref{eq_r2l}), 
temperature, mass density, mass fraction of helium, mass fraction of metals (H$_2$O), 
particle fraction of H$_2$ molecules with respect to the H subsystem, particle fraction of neutral He (column 9) 
and of singly ionized He (column 10) with respect to the He subsystem, and columns 11-13: 
the dimensionless figure functions $s_2$, $s_4$, $s_6$.
}
\end{deluxetable}

\clearpage

%%%%%%%%%%%%%%%%%%%%%%%%%%%%%%%%%%%%%%%%%%%%%%%%%%%%%%%%%%%%%%
\section{Evaluation of the new results}\label{chap_discussion}

In this chapter we discuss to what extent the new results for $M_c$, $Z_{\rm met}$, $Z_{\rm mol}$, 
and $P_{\rm m}$ obtained by using LM-REOS are in agreement with experimental EOS data 
(e.g.\ principal Hugoniot curve), evolution theory, element abundances, and H/He phase separation.

%%%%%%%%%%%%%%%%%%%%%%%%%%%%
\subsection{Core mass $M_c$}

Saumon \& Guillot~\citeyearpar{SG04} found that H-EOS with a small maximum compression ratio of 
only 4 along the Hugoniot curve can yield small core masses lower than 3~M$_{\oplus}$ and comment 
that the apparent relation between the stiffness along the principal Hugoniot and the core mass may 
be not unique. In the same sense, our H-REOS reproduces the maximum compression ratio of 4.25 as 
derived from shock-wave experiments, but the core masses range up to 7~$M_{\oplus}$. 

To understand the indirect effect of the compressibility $\kappa$ of hydrogen on the core 
mass, we study in particular its impact in the deep interior ($\kappa_{\rm met}$), around the 
layer boundary between the envelopes ($\kappa_{\rm m}$), and in the outer molecular region 
($\kappa_{\rm mol}$). The core mass depends essentially and directly on the mass density 
$\rho_{\rm met}$ in the deep interior: The higher $\rho_{\rm met}$, the lower $M_{\rm c}$. Clearly, 
$\rho_{\rm met}$ can either be enhanced by $\kappa_{\rm met}$ or by $Z_{\rm met}$. For example, 
the case of a smaller $\kappa_{\rm met}$ leading to a smaller $\rho_{\rm met}$ occurs with LM-REOS. 
Furthermore, $Z_{\rm met}$ is diminished by both $\kappa_{\rm m}$, which reduces the need for metals 
in order to adjust $J_2$, and the $Z_{\rm mol}$ chosen to reproduce $J_4$. Finally, $Z_{\rm mol}$ 
decreases with $\kappa_{\rm mol}$, for instance in case of SCvH-ppt. Due to this propagation of effects,  
the behavior found in~\citet{SG04} may correspond to the coincidence of a small $\kappa_{\rm m}$ 
and a large $\kappa_{\rm met}$ between 5-15~Mbar, see Fig.~1 in~\citet{SG04}. In agreement with 
\citet{SG04} we conclude that the compressibility along the principal Hugoniot curve, which is 
restricted to densities below about 1~g~cm$^{-3}$, does not determine the core mass alone. 
Experimental data for the hydrogen EOS off the principal Hugoniot curve, i.e.\ near the isentrope, 
are in this context urgently needed. For this, new experimental techniques such as reveberating shock 
waves or precompressed targets can be applied. 

%%%%%%%%%%%%%%%%%%%%%%%%%%%%%%%%%%%%%%%%%%%%%%
\subsection{Abundance of metals $Z_{\rm met}$}

In our new Jupiter models with LM-REOS, $Z_{\rm met}$ is enriched over solar abundance by a factor 
of 5 to 10 and exceeds $Z_{\rm mol}$ by a factor of 4 to 30. This feature is consistent with the 
standard giant planet formation scenario, the core accretion model~\citep{Alibert05,Pollack96}, 
where the planet grows first by accretion of planetesimals onto a solid core embryo. If the 
core has grown such that surrounding nebula gas is attracted, an envelope forms and the planet 
grows by accretion of both gas and planetesimals, either sinking towards the core~\citep{Pollack96} 
or dissolving in the envelope. If the luminosity of the envelopes has reached a critical value, 
the energy loss due to radiation cannot be supplied anymore by infalling planetesimals, and the 
whole planet starts to contract with an even enhanced gas accretion rate (run-away growth). 
Depending on a variety of parameters described in detail in~\citet{Alibert05}, at the end of the 
lifetime of a protoplanetary disk a Jupiter-mass giant planet may have formed with a core mass 
of several $M_{\oplus}$ and a total mass of heavy elements of 30-50~$M_{\oplus}$. Further 
evolution of the planet may include erosion of core material \citep{ChabBar07,Gui03} and homogeneous 
redistribution of destroyed planetesimals due to convection~\citep{Ker04-1}. Both of these 
processes depend on the ability of convection to overcome compositional gradients and their 
efficiency is not known in detail up to now~\citep{ChabBar07,Gui03}. Thus, in the framework of the 
core accretion model, the distribution of heavy elements in our Jupiter models can qualitatively 
be explained as a consequence of planetesimal deliveries in the deep interior and their 
redistribution within the metallic layer by convection.

%%%%%%%%%%%%%%%%%%%%%%%%%%%%%%%%%%%%%%%%%%%%%%
\subsection{Abundance of metals $Z_{\rm mol}$}

With the exception of oxygen, measurements of the abundance of the elements C, N, S, P and the 
noble gases beside He and Ne give two- to fourfold enrichment over solar abundance (see~\citet{Guillot05} for an overview). On the other hand, the Galileo probe gave an O/H ratio of only $0.2\times Z_{\odot}$. It has been argued that the probe fell into a non-representative dry region and the O abundance was still rising with depth when the probe stopped working \citep{Ker04-1}. Our result of $Z_{\rm mol}=1-1.8\times Z_{\odot}$ is too low by a factor of two. We note that LM-REOS based three-layer Jupiter models are difficult to reconcile with the observed atmospheric abundances of heavy elements.

%%%%%%%%%%%%%%%%%%%%%%%%%%%%%%%%%%%%%%%
\subsection{Layer boundary $P_{\rm m}$}

None of the mechanisms known to affect $Z_{\rm mol}$, e.g.\ H/He phase separation or the properties of 
the nebula in the neighbourhood of the forming planet, could explain a depletion of heavy elements 
from the gas phase below solar abundance. To let the Jupiter models have $Z_{\rm mol}>Z_{\odot}$ 
requires $P_{\rm m}>3$~Mbar using LM-REOS. In the three-layer model the discontinuity of the fraction of metals is assumed to occur at the same pressure as the discontinuity of the fraction of helium. 
For helium, the discontinuity has been argued to coincide with a PPT of hydrogen or with the onset 
of H/He phase separation~\citep{SS77-2}. Our QMD simulations show no indications of a PPT so far. 

H/He phase separation on the other hand is a consequence of non-linear mixing and characterized by 
a line $T(x,P)$ of demixing which denotes for a certain concentration $x$ the maximal, pressure-dependent temperature for which the system shows phase separation. While neutral He atoms are expected to separate from metallic hydrogen, ionized helium becomes miscible again at high pressures. However, only few theoretical results for the demixing line exist up to now. Latest simulations~\citep{Pfaff} have a line of demixing at temperatures well below the adiabatic temperature profiles of Jupiter and even of Saturn. Some theoretical EOS of H/He mixtures predict that the Jupiter isentrope intersects the high pressure boundary of the demixing line between 1.5~\citep{FortHubb04} and 7~Mbar~\citep{SS77-1}. At smaller and at higher pressures along the isentrope, the system would mix again. These results are not confident enough to support or exclude a layer boundary around 4~Mbar.

%%%%%%%%%%%%%%%%%%%%%%%%%%%%%%%%%%
\subsection{Possible improvements}

Three-layer models of Jupiter based on QMD data are consistent with the core accretion model, 
but cannot reproduce observed heavy element enrichments by a factor of 2 to 4. Transition pressures 
below 3~Mbar are even accompanied by $Z_{\rm mol}<1\,Z_{\odot}$. Considering the agreement of our   
ab initio QMD results with measured Hugoniot curves and reflectivities for hydrogen, the 
present results for Jupiter point to the importance of three topics to be addressed in the future: 
(i) inclusion of mixing effects in the EOS instead of applying linear mixing, 
(ii) recalculation of the H/He demixing line within an ab initio approach; and 
(iii) adjustment of the structure model in order to account for phase boundaries of H/He demixing 
and He enrichment at deeper levels while still keeping the model as simple as possible.

First results on the effect of nonlinear mixing to the pressure-density relation were obtained by~\citet{Vorberger06} within DFT-MD simulations. They find that the volume at constant pressure is enhanced up to 5\% at temperatures typical for the deep molecular layer where our H-EOS exhibits the highest compressibility, i.e.\ the smallest volume relative to SCvH-i. Taking into account the small mass fraction of metals, a compensation of a 2\% density reduction in average of the H/He component in the outer envelope would require the fraction of metals to be doubled to about 3~$Z_{\odot}$ which is just the average of observed particle species. This effect 
is, therefore, a serious candidate to remove the mismatch between calculated and observed $Z_{\rm mol}$, see discussion above.

Using LM-REOS, the NMT in Jupiter occurs at 0.5 Mbar and is clearly separated from the layer boundary around 4 Mbar with about 80 M$_{\oplus}$ in between. If the order of the latter one will be confirmed by future work on the H/He phase diagram, the possibility of an extended inhomogenous layer enhancing the compressibility should be taken into account in the structure model. 

In order to complement these and former investigations~\citep[e.g.][]{Hub69,SG04} of the interplay between the internal structure of giant planets and the EOS applied, a calculation of the evolution history including H/He phase separation is also necessary.

%%%%%%%%%%%%%%%%%%%%%%%%%%%%%%%%%%%%%
\subsection{Impact for other planets}

Corresponding results can be obtained also for Saturn, Uranus, and Neptune but with higher uncertainty regarding $M_c$, $Z_{\rm mol}$ and $Z_{\rm met}$. For most of the extrasolar giant planets (EGP) detected so far the only structural parameter constrained at all is the minimum mass. Transit detections also allow to determine the mean radius with an error of about 10\%~\citep{Sato05} resulting mainly from the uncertainty in the stellar radius. Even the few transiting planets detected so far have revealed a huge variety of mass-radius relationships. To explain them and to derive implications for their formation and contraction history, well-grounded knowledge of the Solar system giant planets is crucial. For instance, characteristics as helium depletion and compositional gradients are expected to apply also to EGPs \citep{ChabBar07,FortHubb04,Sato05}. 

%%%%%%%%b%%%%%%%%%%%%%%%%%%%%%%%%%%%%%%%%%
\section{Conclusions}\label{chap_summary} 

We have computed three-layer structure models of Jupiter using new EOS data for H, He, and H$_2$O. They are composed of ab initio data in the warm dense matter region covering at least 97\% of Jupiter's total mass and of chemical model EOS at low densities. With 0 to 7~$M_{\oplus}$, the core mass compares with previous results satisfying the same observational constraints. Most of the other properties of QMD-based models such as
(i) up to tenfold enrichment with heavy elements in the metallic layer, 
(ii) small heavy element abundance in the outer envelope, and 
(iii) a transition pressure of at least 3~Mbar, 
are a consequence of the high compressibility of the hydrogen EOS between 0.1 and 10~Mbar along the isentrope, but not along the Hugoniot. Transition pressures of 3 to 5~Mbar agree with estimates of the onset of remixing of hydrogen and helium that are demixing at smaller pressures.

These results clearly underline the importance of calculating the H/He phase diagram with respect 
to the EOS and the region of helium immiscibility in order to improve our understanding of the 
internal structure of hydrogen-rich giant planets. These issues will be addressed in future work.

%%%%%%%%%%%%%%%%
\acknowledgments

We thank J. K{\"o}hn, W.\ Lorenzen, T.R.\ Mattsson, I.\ Iosilevskiy, and T.~Guillot 
for stimulating discussions and D.~Saumon, G.~Chabrier, and 
G.~Kerley for providing us with their EOS data. 
The suggested corrections by the anonymous referee helped to improve the paper.
This work was supported by the Deutsche Forschungsgemeinschaft within the 
Graduiertenkolleg GRK~567, the Sonderforschungsbereich SFB~652, and by the 
High Performance Computing Center North (HLRN) under contract mvp00006. 
We thank the Computing Center of the University of Rostock for assistance.

%%%%%%%%%%%%%%%%%%%%%%%%%%%

\clearpage

\end{document}